\documentclass[fleqn,usenatbib]{mnras}

\usepackage[T1]{fontenc}

\DeclareRobustCommand{\VAN}[3]{#2}
\let\VANthebibliography\thebibliography
\def\thebibliography{\DeclareRobustCommand{\VAN}[3]{##3}\VANthebibliography}

\usepackage{graphicx}	
\usepackage{amsmath}	
\usepackage{amssymb}	

\usepackage{newtxtext,newtxmath}

\title[Analysis of symbiotic candidates]{Spectroscopic and photometric analysis of symbiotic candidates - I. Ten candidates on classical symbiotic stars}

\author[J. Merc et al.]{
J. Merc,$^{1,2}$\thanks{E-mail: jaroslav.merc@student.upjs.sk}
R.~G\'alis,$^{2}$
M. Wolf,$^{1}$
P. Velez,$^{3}$
C. Buil,$^{3}$
F. Sims,$^{3}$
T. Bohlsen,$^{3}$
M.~Vra\v{s}\v{t}\'{a}k,$^{4}$
\newauthor
C. Boussin,$^{3}$
H. Boussier,$^{3}$
P. Cazzato,$^{3}$
I. Diarrasouba,$^{3}$
and F. Teyssier$^{3}$
\\
$^{1}$Astronomical Institute, Faculty of Mathematics and Physics, Charles University, V Hole\v{s}ovi\v{c}k{\'a}ch 2, 180 00 Prague, Czechia\\
$^{2}$Institute of Physics, Faculty of Science, P. J. \v{S}af{\'a}rik University, Park Angelinum 9, 040 01 Ko\v{s}ice, Slovakia\\
$^{3}$Astronomical Ring for Amateur Spectroscopy Group\\
$^{4}$Private observatory Liptovsk\'{a} \v{S}tiavnica, K\v{l}u\v{c}iny 457/74, Slovakia
}

\date{Accepted 2021 July 12. Received 2021 July 12; in original form 2021 June 11}

\pubyear{2021}

\begin{document}
\label{firstpage}
\pagerange{\pageref{firstpage}--\pageref{lastpage}}
\maketitle

\begin{abstract}
Symbiotic stars belong to a group of interacting binaries that display a wide variety of phenomena, including prominent outbursts connected with mass transfer, as well as stellar winds, jets, eclipses, or intrinsic variability of the components. Dozens of new symbiotic stars and candidates have been discovered in recent years. However, there are many objects which are still poorly studied. Some symbiotic candidates suspected in the literature have never been studied spectroscopically. In this contribution, we present the first results of the ongoing campaign focused on symbiotic candidates. In the first paper in the series, we study the nature of ten candidate classical symbiotic stars suspected based on their photometric behaviour, colours or abundance pattern. To confirm or reject the symbiotic nature of the studied candidates, we obtained new spectra and analysed them in detail together with available multi-frequency photometric and spectroscopic observations of the objects. Hen 3-860 and V2204 Oph are genuine symbiotic systems showing typical spectral features of burning symbiotic stars and outbursts in the last 100 years. The first object belongs to the uncommon group of eclipsing symbiotic stars. V1988~Sgr cannot be classified as a genuine burning symbiotic star, but the scenario of an accreting-only symbiotic system cannot be ruled out. Hen 4-204 might be a~bona-fide symbiotic star due to its similarity with the known symbiotic binary BD~Cam. Six other symbiotic candidates (V562~Lyr, IRAS~19050+0001, EC~19249-7343, V1017~Cyg, PN~K1-6, V379~Peg) are either single dwarf or giant stars or non-symbiotic binaries.
\end{abstract}




\section{Introduction}
Symbiotic binaries are unique astrophysical laboratories in the study of stellar evolution, mass transfer, accretion processes, stellar winds, jets, dust formation, or thermonuclear outbursts. They are strongly interacting binaries which consist of a cool giant of spectral type M (less commonly K) and a hot compact star, mostly a white dwarf \citep[see, e.g., reviews by][]{2012BaltA..21....5M,2019arXiv190901389M}. Their orbital periods range from hundreds to thousands of days \citep[e.g.][]{1999A&AS..137..473M}. The mass transfer between the components takes place via the Roche-lobe overflow or via the stellar wind of the cool giant \citep[e.g.][]{1984Ap&SS..99..101A, 1984ApJ...279..252K}, which is also the source of the dense circumbinary envelope of these systems.

Their spectra are a superposition of three components of radiation - two stellar and one nebular \citep[e.g.][]{2005A&A...440..995S}. Usually, the cool giant dominates the spectrum at longer wavelengths (in~IR), and the hot component mainly radiates in the UV and blue part of the optical region. Optical spectra are often rich in emission lines (Balmer lines of \mbox{H}, neutral and ionized \mbox{He}, \mbox{Fe} lines, including forbidden ones). Spectroscopic observations in this wavelength region are usually employed to confirm the symbiotic classification of an object \citep[especially in the case of so-called burning symbiotic stars; see, e.g.][]{2019arXiv190901389M}.

\subsection{Classification criteria}\label{sec:criteria}
Although the exact definition of symbiotic stars has changed over time, the following criteria proposed by \citet{1986syst.book.....K} and later slightly updated by \citet{2000A&AS..146..407B} are most widely used nowadays:
\begin{itemize}
    \item the presence of the absorption features of a late-type giant (e.g. molecular bands of TiO, CN, CO, VO, H$_2$O, and absorption lines of \ion{Ca}{i}, \ion{Ca}{ii}, \ion{Na}{i}, \ion{Fe}{i}, etc.),
    \item the presence of strong \ion{H}{i} and \ion{He}{i} emission lines and either emission lines with an ionisation potential of at least 35 eV and an equivalent width exceeding 1\AA, or an A or F-type continuum with absorption lines of \ion{H}{i}, \ion{He}{i}, and singly ionized metals (for a~symbiotic star in an outburst),
    \item the presence of the Raman-scattered \ion{O}{vi} lines, which is a~sufficient condition, even if the presence of the cool giant is not evident in the spectra).
\end{itemize}

The Raman-scattered \ion{O}{vi} lines at 6825 and 7082 \AA~are observed exclusively in symbiotic binaries \citep{2019ApJS..240...21A}, as the Raman scattering of the highly ionised \ion{O}{vi} 1032 and 1038\,\AA\,\, resonance lines requires a hot source capable of ionising oxygen five times and at the same time sufficient amount of neutral hydrogen in the vicinity \citep{1989A&A...211L..31S}. However, we should note that these lines are observed only in about half of the confirmed symbiotic systems \citep[e.g., the recent census by][]{2019ApJS..240...21A}.

The criteria based on the features in the optical spectrum are used for classification of the so-called burning symbiotic stars. For them, strong emission lines are typical as a consequence of the nuclear burning of hydrogen-rich material on the surface of the hot and luminous white dwarf. These also emit super-soft X-rays. In contrast, accreting-only symbiotic stars do not meet the above criteria. They do not manifest prominent optical emission lines, but are bright in UV, show significant flickering in this spectral region and are hard X-ray sources \citep[e.g.][]{2013A&A...559A...6L,2016MNRAS.461L...1M,2021arXiv210402686M}.

In addition to the overall appearance of optical spectra, other indicators are often used to identify promising symbiotic candidates. They are often suspected based on the peculiar photometric behaviour (outbursts, typical quasi-periodic quiescent variations, etc.). The position in the near IR 2MASS colour-colour diagram, where symbiotic stars occupy a distinct region, can also indicate the symbiotic nature of the object \citep[e.g.][]{2007MNRAS.376.1120P,2008A&A...480..409C,2014A&A...567A..49R}. However, the same region is also shared with other groups of objects, such as planetary nebulae, Be stars, or T~Tauri stars (see Fig. \ref{fig:ir}), which can mimic the symbiotic appearance or behaviour. Recently, \citet[][]{2019MNRAS.483.5077A} proposed a set of IR criteria for classifying the symbiotic stars based on the IR photometry from 2MASS and WISE.

\citet{2017A&A...606A.110I} presented spectroscopic criteria based on the ratio of emission line fluxes to distinguish between planetary nebulae and symbiotic stars, extending the diagnostic diagrams of \citet{1995PASP..107..462G} which are applicable only if [\ion{O}{iii}] emission lines are observed together with Balmer lines. In our recent work \citep{2020MNRAS.499.2116M}, we have proposed to employ the position in the \textit{Gaia} DR2 HR diagram, or the unusual X-ray-to-optical flux ratios as criteria to discard disputable symbiotic candidates and, to a~lesser extent, to identify promising systems.

Ongoing systematic searches for symbiotics in the Milky Way \citep[e.g.][]{2014MNRAS.440.1410M} and in the nearby galaxies \citep[e.g.][]{2018arXiv181106696I} have resulted in the number of known symbiotic systems to grow rapidly. Until 2019, the last catalogue was published by \citet{2000A&AS..146..407B}. \citet{2019ApJS..240...21A} published a new census of galactic and extragalactic symbiotic stars with more than twice as many objects in comparison with the previous catalogue.

Based on the recent progress in the study of symbiotics, we have decided to prepare a new database of these interacting binaries \citep[New Online Database of Symbiotic Variables; ][]{2019RNAAS...3...28M}. The purpose of the database is not only to serve as a catalogue of data, but we have also prepared a web portal\footnote{http://astronomy.science.upjs.sk/symbiotics/} for easy access to this information. The current version of the database consists of more than 400 confirmed and suspected galactic symbiotics as well as 74 confirmed and 83 suspected extragalactic symbiotic systems which are located in 14 galaxies.

\begin{table}
\caption{List of objects analysed in this paper.}
\label{table:targets}
\centering
\begin{tabular}{rlrr}
\hline\hline
No. & Name & $\alpha_{2000}$ [h m s] & $\delta_{2000}$ [d m s]\\\hline
1&Hen 3-860 & 13 06 12.916 & -53 22 52.50 \\
2&V2204 Oph & 18   26 01.923 & +11 55 09.53 \\
3&V1988 Sgr & 18 27 57.308 & -27 37 23.11 \\
4&V562 Lyr & 18 31 13.817 & +46 58 34.67 \\
5&IRAS 19050+0001 & 19   07 37.430 & +00 06 09.10 \\
6&EC 19249-7343 & 19 31 02.317 & -73 37 07.64 \\
7&V1017 Cyg & 19 56 15.815 & +53 19 12.12 \\
8&PN K1-6 & 20 04 14.278 & +74 25 35.93 \\
9&Hen 4-204 & 22 45 01.416 & -44 52 38.52 \\
10&V379 Peg & 23 53 50.821 & +23 09 18.09\\\hline
\end{tabular}
\end{table}

Many of the known symbiotic stars are only poorly studied. Moreover, there are several objects proposed to be symbiotic stars based only on their photometric appearance or behaviour. In this first paper in the series, we have selected ten symbiotic candidates from our New Online Database of Symbiotic Variables for which no or very limited spectroscopic information is available. The selected objects with coordinates are listed in Table \ref{table:targets}. The paper is organised as follows: in Section \ref{sec:observations} we describe the spectroscopic and photometric observational data considered and in Section \ref{sec:results} we discuss the results for the selected symbiotic candidates.

\section{Observational data}\label{sec:observations}
\begin{figure*}
\centering
\includegraphics[width=0.78\textwidth]{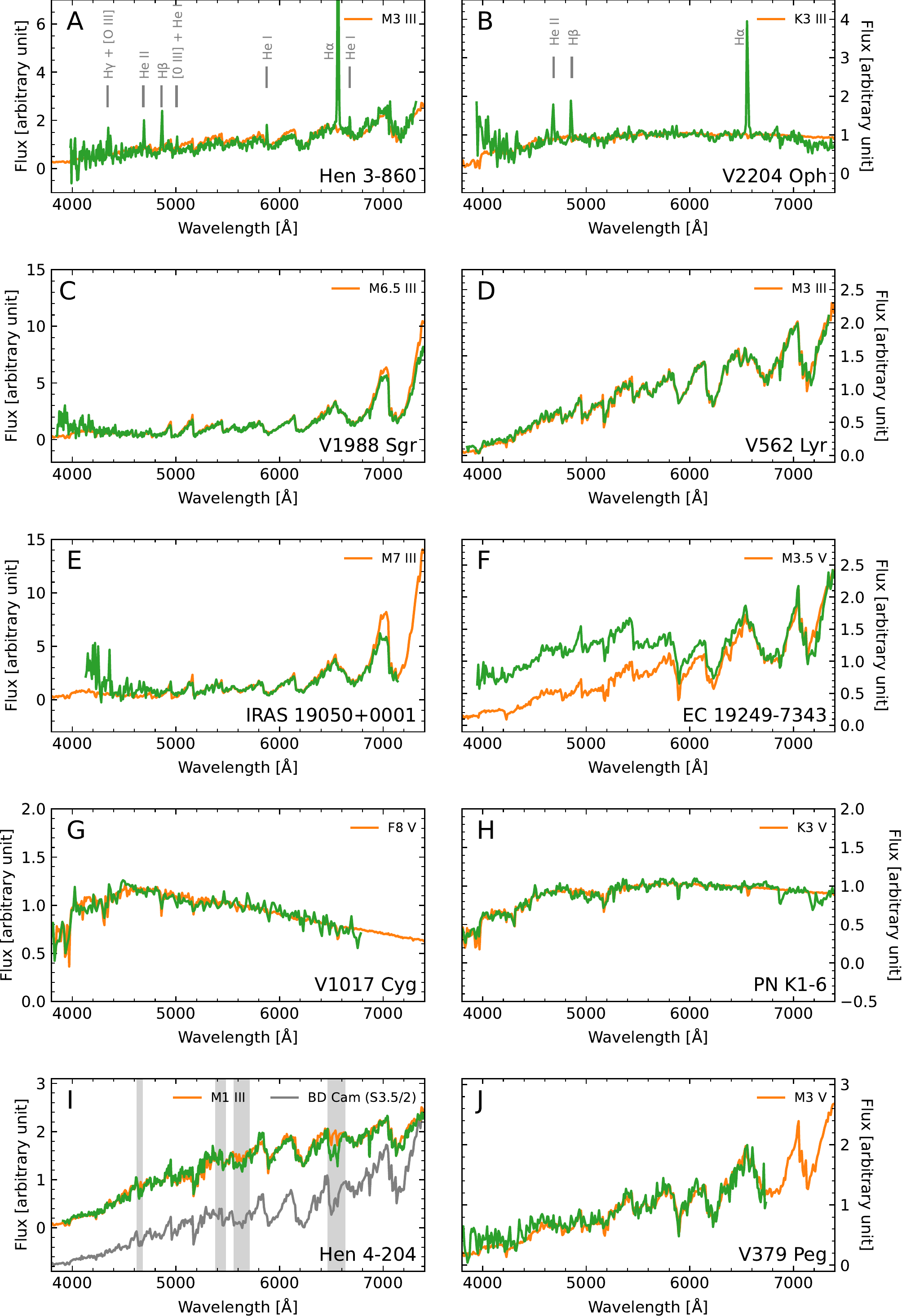}
\caption{Spectra of the studied symbiotic candidates. The spectra obtained in the scope of our campaign and the best-fitting empirical spectra \citep{2011A&A...532A..95F,2017ApJS..230...16K} are shown in green and orange, respectively. In the panels A and B, the identification of the most prominent emission lines is depicted. In panel I, the ARAS spectrum of BD Cam, the symbiotic star with S-type star component is shown in grey. The position of the most prominent ZrO bands is depicted by shaded areas. All spectra were de-reddened by the $E_{\rm (B-V)}$ values listed in Table \ref{table:photometric_data} using the reddening law of \citet{1989ApJ...345..245C} and adopting the total to selective absorption ratio $R = 3.1$.}
\label{fig:spectra}
\end{figure*}

To confirm or reject the symbiotic nature of the studied candidates, we obtained new spectroscopic observations by organizing an international observing campaign in cooperation with the ARAS Group\footnote{https://aras-database.github.io/database/symbiotics.html} \citep[\textit{Astronomical Ring for Amateur Spectroscopy};][]{2019CoSka..49..217T}. ARAS is an initiative dedicated to the promotion of amateur astronomical spectroscopy and pro/am collaboration. The network consists of observers equipped with small telescopes (20 to 60 cm) with spectrographs of different resolution (500 to 15\,000). The obtained spectra (Fig. \ref{fig:spectra}) were compared to the ones from empirical libraries of stellar spectra of \citet{2011A&A...532A..95F} (MILES; giants) and \citet{2017ApJS..230...16K} (dwarfs, giants earlier than M0). The spectra were down-sampled to the resolution of the ARAS data in order to obtain the spectral classification of the studied objects. If present, emission spectral lines have been identified in each spectrum.

\begin{figure*}
\centering
\includegraphics[width=0.78\textwidth]{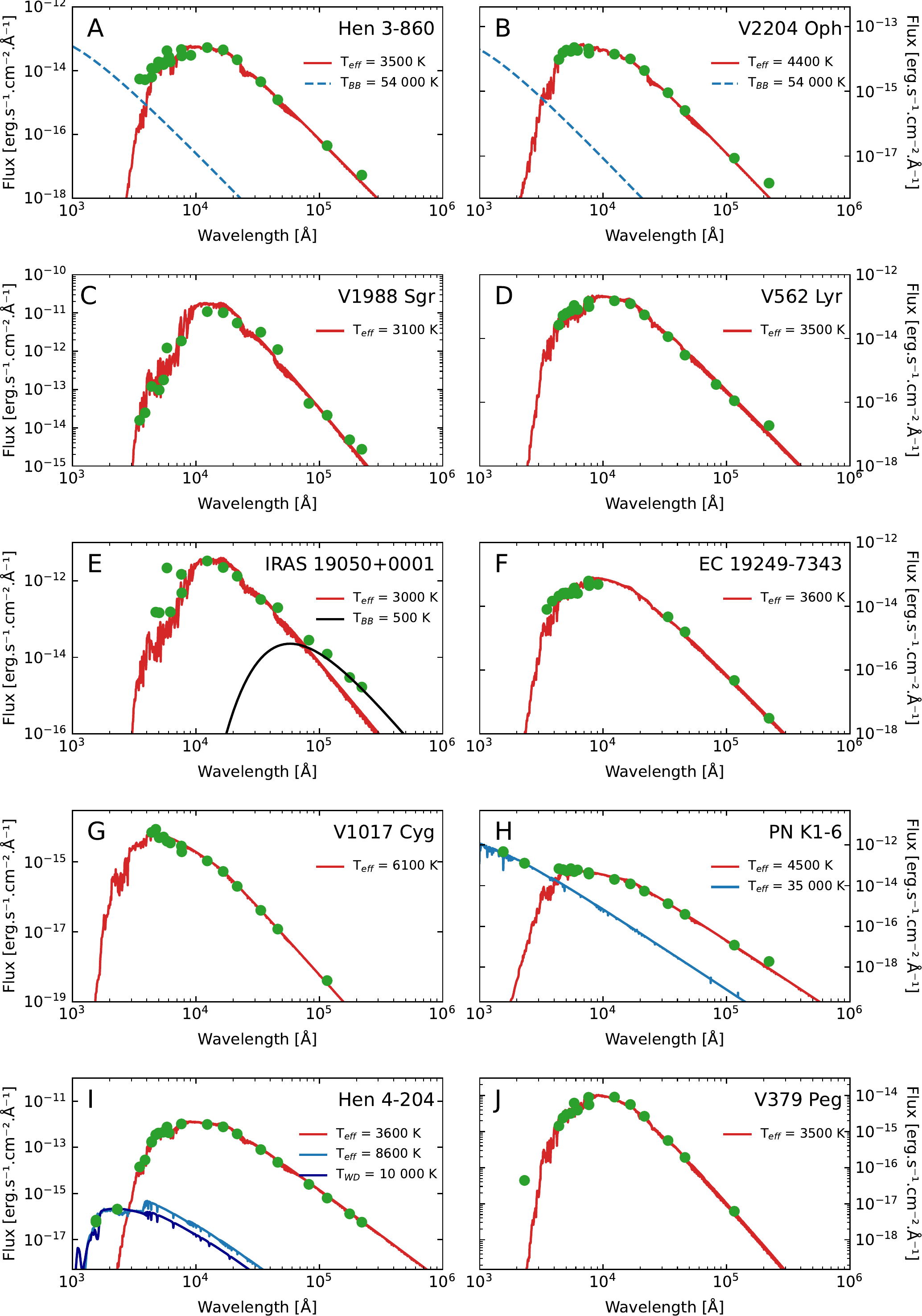}
\caption{Multi-frequency SEDs of the studied symbiotic candidates. Measurements from GALEX, SkyMapper, APASS, Gaia EDR3, 2MASS, AKARI, and WISE are shown in green. The best fitting theoretical spectra \citep{2014IAUS..299..271A} are plotted in red. If two sources of radiation are present, the hotter is shown in blue (stellar spectra for PN K1-6 and Hen 4-204, black body of 54\,000\,K for Hen 3-860 and V2204 Oph; see the text for details). In the case of Hen 4-204, we also show the best fitting WD spectrum in dark blue \citep[][]{2010MmSAI..81..921K}. In the case of IRAS 19050+0001, the excess in IR presumably caused by a dust is modelled by the black body (shown in black). The fluxes were de-reddened by $E_{\rm (B-V)}$ values listed in Table \ref{table:photometric_data} using reddening law of \citet{1989ApJ...345..245C} and adopting the total to selective absorption ratio $R = 3.1$.}
\label{fig:seds}
\end{figure*}

\begin{figure}
\centering
\includegraphics[width=\columnwidth]{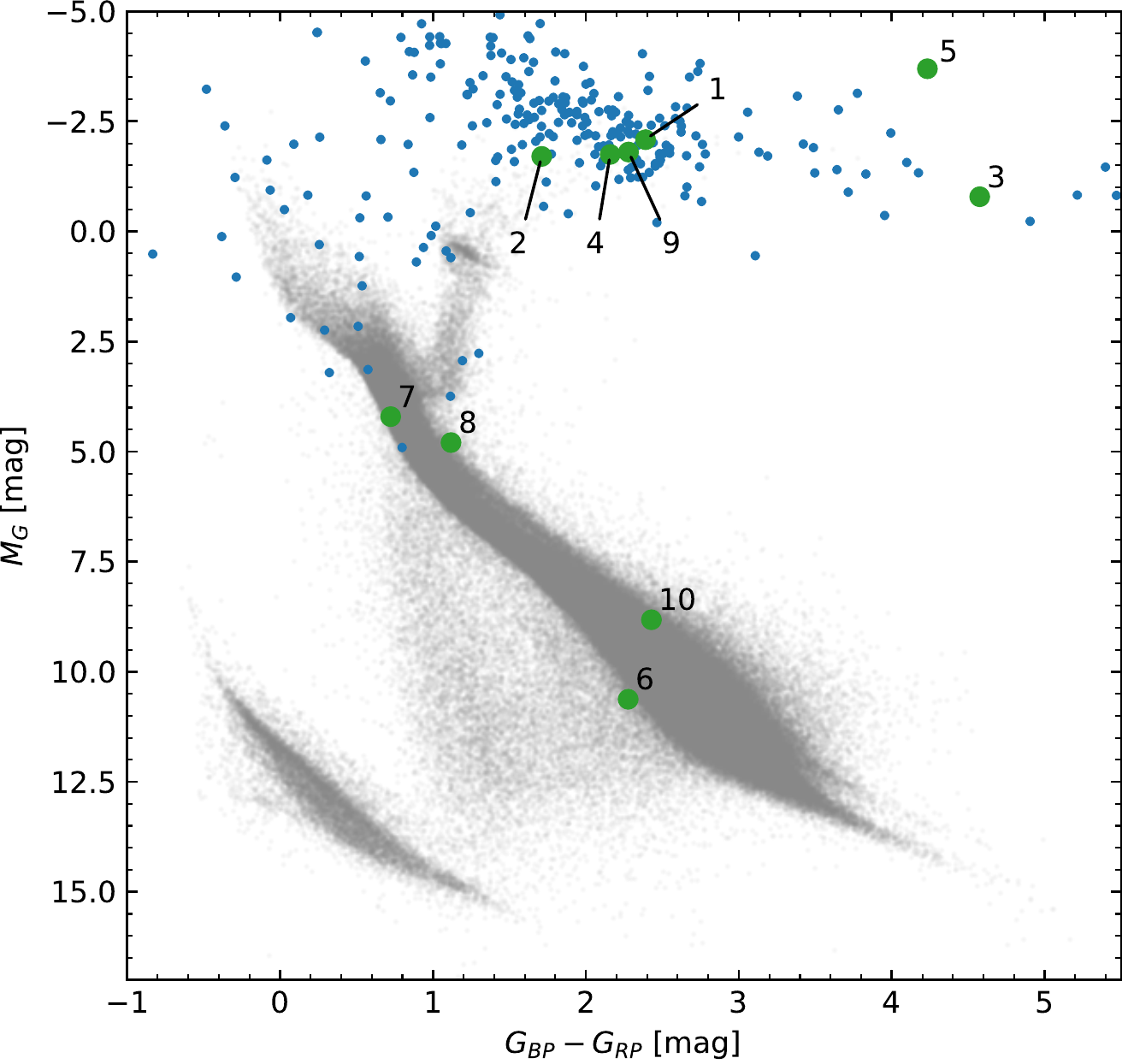}
\caption{Position of studied objects (green symbols; labels according to Table \ref{table:targets}) in the \textit{Gaia} HR diagram of a sample of stars within 200\,pc with reliable astrometry \citep{2018A&A...616A..10G}. Known symbiotic variables from the New Online Database of Symbiotic Variables \citep{2019RNAAS...3...28M} are plotted with blue symbols.}
\label{fig:hr}
\end{figure}

\begin{figure}
\centering
\includegraphics[width=0.98\columnwidth]{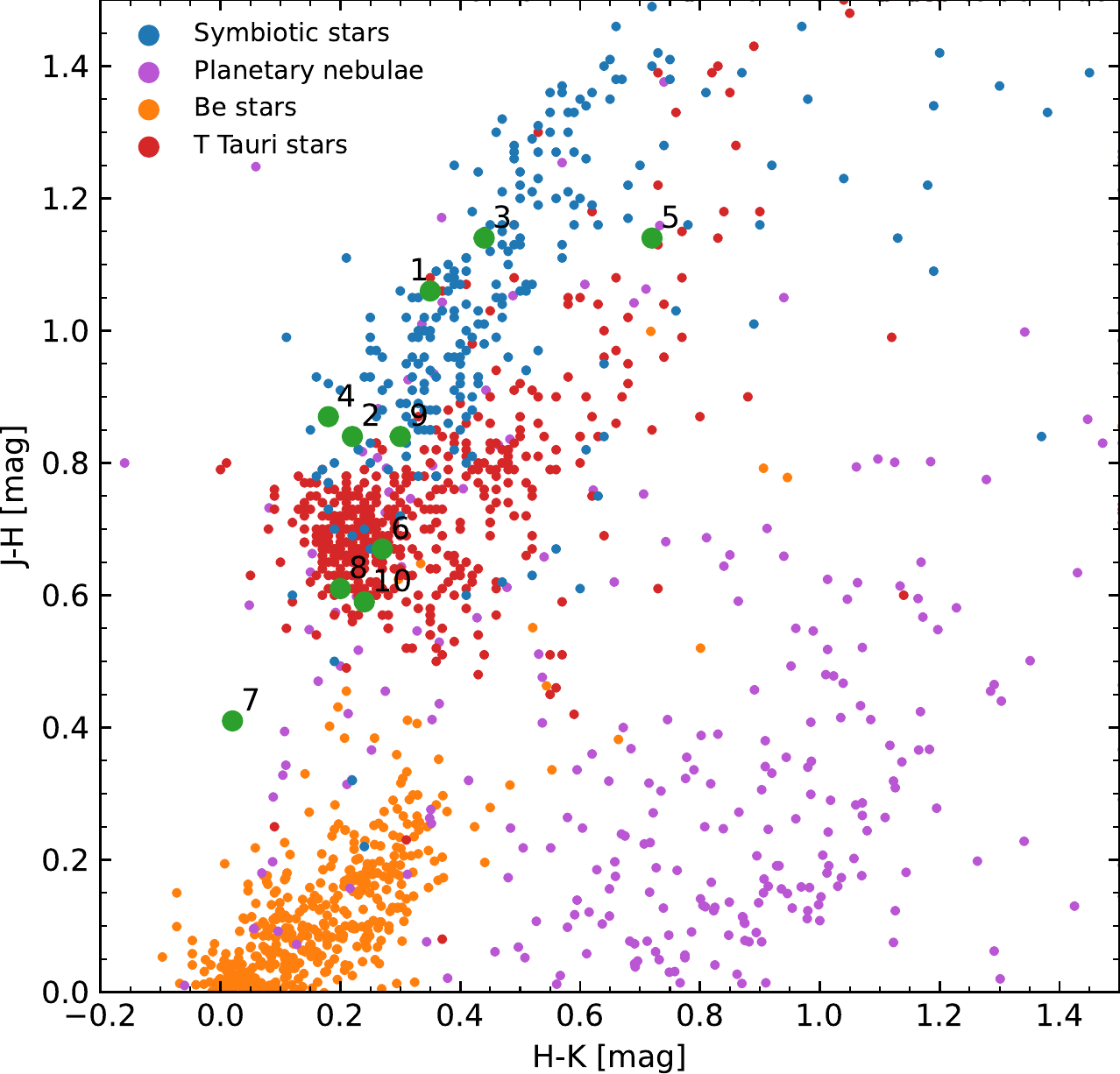}
\caption{Position of studied objects (green symbols; labels according to Table \ref{table:targets}) in the near IR colour-colour diagram based on 2MASS observations \citep{2006AJ....131.1163S}. The positions of the known symbiotic stars, planetary nebulae \citep{2005MNRAS.357..732R}, Be stars \citep{2005NewA...10..325Z}, and T~Tauri stars \citep{2005AJ....129..829D} are shown with blue, violet, orange and red symbols, respectively.}
\label{fig:ir}
\end{figure}

For the classification of the symbiotic candidates, we have retrieved all available information on the objects from the literature and supplemented them with the data from Gaia EDR3 \citep{2020arXiv201201533G}, GALEX \citep{2017ApJS..230...24B}, SkyMapper \citep{2018PASA...35...10W}, APASS \citep{2015AAS...22533616H}, 2MASS \citep{2006AJ....131.1163S}, AKARI \citep{2010A&A...514A...1I}, and WISE \citep{2010AJ....140.1868W}. From this data, we have also constructed the multi-frequency spectral energy distributions (SEDs; Fig. \ref{fig:seds}), colour-magnitude diagram (Fig. \ref{fig:hr}), and near-infrared (NIR) colour-colour diagram (Fig. \ref{fig:ir}). The constructed SEDs of the symbiotic candidates were compared with the BT-Settl grid of theoretical spectra \citep{2014IAUS..299..271A} downloaded from Theoretical spectra webserver at the Spanish Virtual Observatory Theoretical Model Services\footnote{http://svo2.cab.inta-csic.es/theory/newov2/index.php} in order to estimate the parameters of the radiation sources. The spectral types were estimated from the effective temperatures using the statistical relations of \citet{2020RAA....20..139M}.

To study the photometric variability of the selected objects, we have employed the data obtained from the All-Sky Automated Survey \citep[ASAS; $V$ filter;][]{1997AcA....47..467P}, All-Sky Automated Survey for Supernovae \citep[ASAS-SN; \textit{V}~and \textit{g} filters; ][]{2014ApJ...788...48S, 2017PASP..129j4502K}, and the Zwicky Transient Facility (ZTF) survey \citep[\textit{r}~and \textit{g} filters; ][]{2019PASP..131a8003M}. The photometric data were subjected to visual inspection and used for construction of the light curves of the selected candidates (Fig. \ref{fig:LCs}). Period analysis was performed using the Peranso software\footnote{https://www.cbabelgium.com/peranso/} to obtain the information on any periodic behaviour of the  studied symbiotic candidates. The properties of the selected objects are listed in Table \ref{table:photometric_data}.

\begin{table*}
\caption{Properties of the studied objects. Data were obtained from Gaia EDR3 \citep[$G$, $\pi$, $\mu$, $G_{\rm BP}$, $G_{\rm RP}$;][]{2020arXiv201201533G}, GALEX \citep[FUV, NUV;][]{2017ApJS..230...24B}, and 2MASS \citep[$J$, $H$, $K_{\rm S}$;][]{2006AJ....131.1163S}. The values of extinction $E_{\rm (B-V)}$ used in this work (the last column) were taken from \citet{2011ApJ...737..103S} and adjusted for the closest objects (as the catalogue values correspond to the total Galactic extinction in the given direction): (6) EC~19249-7343, (10) V379 Peg. Higher value than given by the reddening map was adopted for (5) IRAS 19050+0001 (see the text for details). Symbol(s) < denote objects which were observed by GALEX but were below the detection limit of the survey.}
\label{table:photometric_data}
\centering
\begin{tabular}{lrrrrrrrrrrr} \hline \hline
No. & $G$ & $\pi$ & $\mu$  & FUV & NUV & $G_{\rm BP}$ & $G_{\rm RP}$ & $J$ & $H$ & $K_{\rm S}$ & $E_{\rm (B-V)}$ \\
& [mag] & [mas] & [mas\,yr$^{-1}$] & [mag] &[mag] &[mag] &[mag] &[mag] &[mag] &[mag]& [mag] \\\hline
1 & 13.19 & 0.10 $\pm$ 0.03 & 8.65 & - & - & 14.91 & 11.94 & 9.79 & 8.73 & 8.38& 0.45 \\
2 & 13.61 & 0.04 $\pm$ 0.02 & 4.13 & - & - & 14.56 & 12.64 & 11.16 & 10.32 & 10.1& 0.16\\
3 & 9.24 & 1.39 $\pm$ 0.14 & 14.54 & - & - & 12.73 & 7.72 & 3.93 & 2.79 & 2.35& 0.34\\
4 & 10.90 & 0.28 $\pm$ 0.01 & 4.77 & < & < & 12.05 & 9.84 & 8.28 & 7.41 & 7.23& 0.04\\
5 & 11.86 & 0.32 $\pm$ 0.12 & 2.61 & - & - & 16.28 & 10.12 & 6.16 & 5.02 & 4.30& 1.50\\
6 & 12.06 & 51.69 $\pm$ 0.09 & 286.52 & - & - & 13.05 & 10.78 & 9.18 & 8.51 & 8.24& 0.00\\
7 & 15.04 & 0.88 $\pm$ 0.02 & 8.22 & - & - & 15.44 & 14.48 & 13.82 & 13.41 & 13.39& 0.19\\
8 & 12.30 & 3.86 $\pm$ 0.03 & 28.19 & 14.00 & 14.67 & 12.86 & 11.51 & 10.62 & 10.01 & 9.81& 0.18\\
9 & 8.82 & 0.73 $\pm$ 0.02 & 5.15 & 22.32 & 20.09 & 10.03 & 7.74 & 6.26 & 5.42 & 5.12& 0.01\\
10 & 14.03 & 9.36 $\pm$ 0.02 & 63.8 & < & 21.67 & 15.36 & 12.89 & 11.35 & 10.76 & 10.52 &0.03\\ \hline
\end{tabular}
\end{table*}

\section{Results and discussion}\label{sec:results}
In this section, we discuss the known details on the selected objects, the results obtained by the analysis of photometric and spectroscopic observations, and the proposed classifications of the studied symbiotic candidates.

\subsection{Hen 3-860} The object was selected based on its peculiar light curve from the ASAS-SN survey \citep{2014ApJ...788...48S,2017PASP..129j4502K} showing recent brightening (2017\,-\,2019) resembling a symbiotic outburst and eclipse-like features. Possible symbiotic classification was first suggested by the amateur astronomer Gabriel Murawski. Previously, the star was classified as H$\alpha$ emitter by \citet{1966PhDT.........3W} and \citet{1976ApJS...30..491H}.

The spectroscopic observations obtained in the scope of our campaign (Fig. \ref{fig:spectra}A) confirmed the symbiotic nature of the object. In addition to the M2-3 III continuum, strong emission lines of \ion{H}{i}, \ion{He}{i}, and \ion{He}{ii} are observed in the spectrum, satisfying the spectroscopic criteria presented in Section \ref{sec:criteria}. The SED (Fig. \ref{fig:seds}A) also suggests an emission excess in the blue optical spectral region. The lower limit of the temperature of the hot component in the system of 54\,000\,K is given by the method of \citet{1994A&A...282..586M} employing the emission line with the maximum ionisation potential observed in the spectrum (i.e., \ion{He}{ii}). The position of the object in the HR diagram (Fig. \ref{fig:hr}) is consistent with other symbiotics. The NIR colours of the star are typical for S-type symbiotic binaries (see its position in Fig. \ref{fig:ir}). Hen 3-860 also satisfies the IR criteria for S-type symbiotic stars proposed by \citet[][]{2019MNRAS.483.5077A,2021MNRAS.502.2513A}.

The object was observed in the outburst in 2017-2019 (Fig. \ref{fig:LCs}A). 
The study of the recent photometry of the system revealed the presence of eclipses. We have used this feature to estimate the orbital period to be $\sim$\,600\,days. That is a typical value for \mbox{S-type} symbiotics \citep[e.g.,][]{2013AcA....63..405G}. All these allow us to conclude that Hen 3-860 is a newly discovered southern symbiotic system, representative of an uncommon group of eclipsing symbiotics. A detailed analysis of this newly discovered system will be presented in the subsequent paper.


\subsection{V2204 Oph} The object has been observed in at least two brightenings, by \citet{1926AJ.....36..122R} and by \citet{1983MitVS...9...87S}. Samus' classified the star as a possible symbiotic binary. No spectrum has been obtained since then.

The continuum spectrum (Fig. \ref{fig:spectra}B) and multi-frequency SED (Fig. \ref{fig:seds}B) of the object is consistent with a K3 giant. The spectral type of the cool component, the presence of the emission lines of \ion{H}{i}, \ion{He}{i}, and the strong line of \ion{He}{ii} 4686\,\AA, together with reported outbursts of the star \citep{1926AJ.....36..122R,1983MitVS...9...87S} allows us to confirm it as a yellow symbiotic binary. A spectrum of higher S/N would be needed to precisely characterise the symbiotic nebula and the hot component of the system. The lower limit of the temperature of the hot component is 54\,000\,K according to the method of \citet[][]{1994A&A...282..586M}. The position of V2204 Oph in the HR diagram (Fig. \ref{fig:hr}) is typical for other symbiotics. In the NIR colour-colour diagram, the object is located in the bottom-left part of region occupied by the S-type symbiotic stars (Fig. \ref{fig:ir}). This very same position in the diagram is occupied by another well-known yellow symbiotic star, AG Dra. The IR properties also satisfies the criteria for S-type symbiotic stars which were proposed by \citet[][]{2019MNRAS.483.5077A,2021MNRAS.502.2513A}.

ASAS-SN and ZTF photometric observations revealed the possible presence of the period $\sim$450\,days, which may be due to the orbital motion. However, the time interval covered by the ZTF observations (shown in Fig. \ref{fig:LCs}B) is too short to be conclusive and there is a rather large scatter in the light curves of V2204 Oph in the ASAS-SN data. This could be because the star is close to the limit of the survey ($V \sim 14.5$\,mag). On the other hand, giants in symbiotic stars often pulsate on timescales of 50 - 200 days, which introduce additional scatter to the quiescent light curves. Only a well-sampled long-term light curve and/or radial velocity measurements would help to confirm the presence of the period and to refine its value.

\subsection{V1988 Sgr} The star is included in the General Catalogue of Variable Stars (GCVS) as a possible symbiotic binary \citep{2017ARep...61...80S} based on photometric variability (possible brightening by <\,1.3\,mag detected on five photographic plates) identified by \citet{1962AJ.....67..228H}. The object is also listed as a suspected symbiotic star in \citet{1986syst.book.....K}. We should note, that GCVS and SIMBAD list IRC-30385 as a cross-identification of V1988~Sgr in the Two-Micron Sky Survey \citep{1969tmss.book.....N}. The star is classified in the survey as M7. However, in the catalogue itself, IRC-30385 is identified as LP Sgr, which is a Mira variable with Me spectrum according to GCVS. No spectrum of either of the two variable stars is available in the literature. We have analysed both objects. The field around V1988~Sgr is shown in the attachment in Fig. \ref{fig:v1988sgr}.

V1988 Sgr is the most reddened of the studied symbiotic candidates (see HR diagram in Fig. \ref{fig:hr}) and its position in the NIR colour-colour diagram is typical for S-type symbiotic binaries (Fig. \ref{fig:ir}). The light curve of the object (Fig. \ref{fig:LCs}C, C', and D) revealed the presence of the long-term variability with the period around 3\,125 days, and the semi-regular pulsations with period $\sim 90$\,days. Such photometric variability is often visible in the light curves of symbiotic stars \citep[e.g.,][]{2013AcA....63..405G}. On the other hand, the obtained spectra of V1988 Sgr (Fig. \ref{fig:spectra}C) are consistent with M6.5III star and do not show any emission lines nor excess in the blue part, which are typical for the burning symbiotic stars. Therefore, these data don't allow us to confirm the symbiotic nature of the system. The SED of the object (Fig. \ref{fig:seds}C) confirms the spectral classification and these data do not indicate the presence of a hot symbiotic component.

It could be that V1988 Sgr is not a burning symbiotic star but an accreting-only one. As mentioned in Section \ref{sec:criteria}, such symbiotics don't show emission lines in the spectrum, they are bright in UV and show significant flickering in this spectral range. Unfortunately, no observations of V1988 Sgr in this part of the spectrum are available. Flickering is often detectable also at optical wavelengths, with the highest amplitudes observed at the shortest wavelenths (e.g., in the Johnson \textit{U} filter). As the observations in the \textit{U} filter are rather infrequent in the era of CCDs, we executed a 90-min observing run at JD\,2\,459\,329.8 (April 25, 2021) in Johnson \textit{B} filter using Danish 1.54-meter telescope at La Silla, Chile. In total, 35 frames were obtained with exposure time of 120\,s each. We have not detected any significant flickering above the observational noise. For this reason, the data do not allow us to confirm the symbiotic status of V1988~Sgr. On the other hand, as \citet{2021arXiv210402686M} pointed out, even in the case of well-known accreting-only symbiotic stars such as SU~Lyn or MWC~560, flickering is not always detectable.

Based on these findings, we cannot classify V1988 Sgr as a~genuine symbiotic, but at the same time, we do not rule out its possible accreting-only symbiotic classification. It is worth noting that its IR properties satisfy the IR criteria proposed by \citet[][]{2019MNRAS.483.5077A,2021MNRAS.502.2513A} for S-type symbiotics, having W1-W2$\geq$0.09, similarly to MWC~560. On the other hand, it remains an open question whether the brightening in 1924 identified by \citet{1962AJ.....67..228H} was a symbiotic outburst, or was a consequence of the periodic photometric variability observed in the star even today. For definitive classification, new observations over time and/or in other spectral regions will be needed.

We have also analysed the data of LP Sgr, a nearby star which is sometimes confused with V1988 Sgr in the literature. The ASAS-SN and ASAS light curves of LP Sgr phased with the period of 219.93\,days are shown in Fig. \ref{fig:LCs}E. These data are consistent with the Mira pulsations with a rather high amplitude of several magnitudes. The spectrum of LP Sgr, which we obtained together with observations of V1988 Sgr, shows M3.5 III continuum and H$\alpha$ in emission. The spectrum was obtained at phase $\sim$\,0.1 ($\varphi$ = 0 being a brightness maximum). The data suggest that LP Sgr is a single Mira pulsator.

\subsection{V562 Lyr} Unusual behaviour of the star was first noted by \citet{2000IBVS.4898....1D}, who detected the object on three plates to be at least 3.3\,mag fainter than usual. The variability was later studied by \citet{2000IBVS.4926....1G}. They confirmed the presence of a single long-lasting deep eclipse (1\,000 - 1\,200 days) or an R CrB variability and suggested that the object might be an eclipsing binary with a very long period, R CrB variable or a symbiotic star. No spectrum was available.

Although the star is generally referred as a confirmed symbiotic binary (e.g. in SIMBAD), the obtained spectrum (Fig. \ref{fig:spectra}D) and SED (Fig. \ref{fig:seds}D) are very well consistent with a single M3 giant and no emission lines are observed. The position of the object in the HR and NIR colour-colour diagrams (Figs. \ref{fig:hr} and \ref{fig:ir}) is consistent with the giant classification. Moreover, the star has not been detected by \textit{GALEX} in UV, which would be expected for a symbiotic star (either burning or accreting-only) or a binary with a hot companion. Analysis of the photometric data from ASAS-SN and ZTF of V562 Lyr shows a possible presence of irregular variability on the timescale $\sim$16 days with an amplitude of <\,0.2\,mag in the $V$ filter. No minima are seen in the recent light curve.

\subsection{IRAS 19050+0001} The highly reddened object is relatively faint in the $V$ filter, but bright in the red spectral region, and especially in IR. The star seems to be included in the lists of symbiotic stars by mistake, as there is a confusion about the position of the object in the literature. Databases such as SIMBAD give NSV 11749, a known symbiotic binary \citep{2012PASP..124.1262B,2014A&A...567A..49R}, at the position of IRAS 19050+0001. This object also appears on the list by \citet{2019ApJS..240...21A}. However, NSV 11749 is another object. For this reason, there is no spectroscopic information on IRAS 19050+0001 in the literature.

Its position in the \textit{Gaia} HR diagram (Fig. \ref{fig:hr}) as well as in the NIR colour-colour diagram (Fig. \ref{fig:ir}) confirmed that the object is very red. Because of the high reddening and consequent faintness of the object in the blue part of the optical spectrum \citep[<\,20\,mag in the $B$ filter;][]{2008AJ....136..735L}, we have used only the part redwards of 5\,500\,\AA\,\,for the spectral classification. To obtain an appropriate fit to any of the empirical spectra, we have increased the extinction value to 1.5 \citep[the value given by the reddening map is $\sim$0.6;][]{2011ApJ...737..103S}. That part of the spectrum is well consistent with an M7 III star (Fig. \ref{fig:spectra}E). The multi-frequency SED (Fig. \ref{fig:seds}E) seems to be more consistent with a slightly cooler star (M8), with the IR excess consistent with the blackbody model with the temperature of $\sim$500\,K.

\begin{figure*}
\centering
\includegraphics[width=0.78\textwidth]{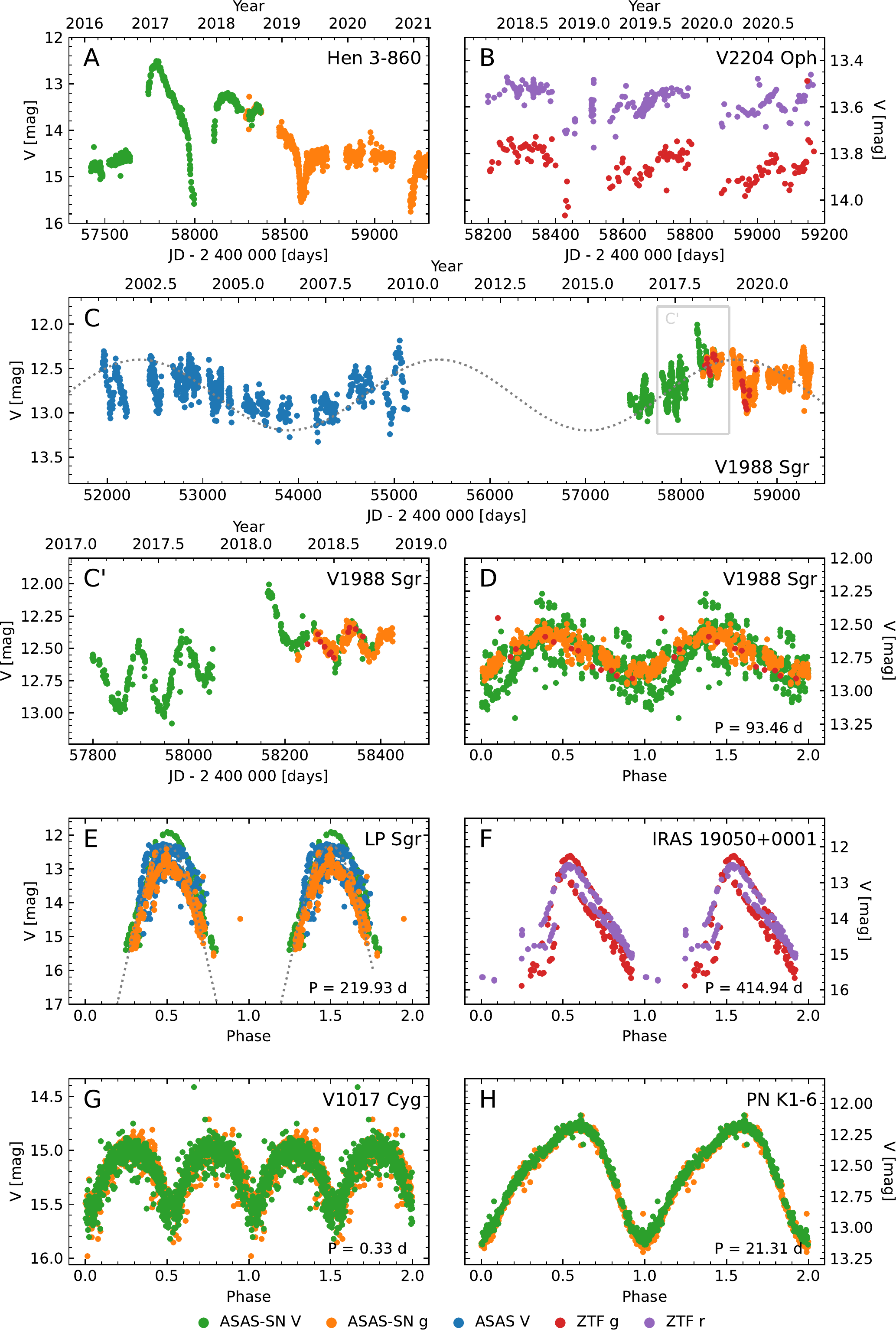}
\caption{Light curves of the studied symbiotic candidates. \textbf{A}: Light curve of Hen 3-860. The $g$ light curve was shifted by -0.66 mag to the level of $V$~magnitudes for clarity. \textbf{B}: Light curves of V2204 Oph. \textbf{C}: Light curve of V1988 Sgr. ASAS $V$ data was shifted by +0.25 mag, ASAS-SN $g$~and ZTF $g$ by -0.95 mag to the level of ASAS-SN $V$ ones. Dotted grey curve shows long-term variation with the period of 3\,125 days. \textbf{C'}: Part of the light curve of V1988 Sgr showing the pulsation features. \textbf{D}: Light curves of V1988 Sgr with subtracted long-term variability, phased with the period of 93.46 days. \textbf{E}: Light curves of LP Sgr, phased with the period of 219.93 days. Data in the $g$ filter were shifted by -0.60 mag to the level of $V$ ones. \textbf{F}: Light curves of IRAS 19050+0001 phased with period of 414.94 days. Data in the $g$ filter were shifted by -3.5 mag to the level of the $r$ filter. \textbf{G}: Light curves of V1017 Cyg phased with period 0.33 days. Data in the $g$ filter were shifted by -0.22 mag to the level of $V$ ones. \textbf{H}:~Light curves of PN K1-6 phased with period of 21.31 days. Data in the $g$ filter were shifted by -0.44 mag to the level of $V$ ones.}
\label{fig:LCs}
\end{figure*}

The position of IRAS 19050+0001 in the NIR colour-colour diagram (Fig. \ref{fig:ir}) suggests that it might be a symbiotic star, possibly of a D-type. However, our spectroscopic observation have not shown the presence of emission lines in the spectrum preventing the classification of the object as a burning symbiotic star. The analysis of the ZTF light curves of the star revealed the presence of the pulsations with the period of 414.94 days. The light curves phased with this period are shown in Fig. \ref{fig:LCs}F. 

Considering the shape of the light curves, the obtained value of the period and the spectral type, we argue that the object is a (probably single) Mira variable. Using the period-luminosity relation of \citet{2008MNRAS.386..313W} for O-rich Miras, we have obtained the distance of 2.6\,kpc for IRAS 19050+0001, which is consistent with the photo-geometric distance of 2.4 (2.0-3.1)\,kpc obtained by \citet{2021AJ....161..147B} using the \textit{Gaia} EDR3 data. The dust produced in the stellar atmosphere  would explain the higher value of the extinction than the one (interstellar) given by the dust map. The presence of the warm dust and the Mira pulsations can also explain the IR excess in the SED of the object and the difference between the spectral type obtained from the spectrum and the SED, respectively.

\subsection{EC 19249-7343} The object is listed as a symbiotic candidate in the catalogue of \citet{2013MNRAS.431..240O} based on the low-resolution spectrum from the Edinburgh-Cape survey. They suggested that the peculiar spectrum with molecular bands is similar to the recurrent symbiotic nova T CrB. The star was also included in the catalogues of M dwarfs \citep{2011AJ....142..138L,2013MNRAS.435.2161F}.

According to the \textit{Gaia} EDR3 data, the object is a high-proper motion star located at a distance of 19.3 pc \citep{2021AJ....161..147B, 2020arXiv201201533G}. At such distance, its brightness is well consistent with a dwarf star (Fig. \ref{fig:hr}), which precludes the symbiotic classification of the object. The  position of EC 19249-7343 in the NIR colour-colour diagram (Fig. \ref{fig:ir}) is also not consistent with the symbiotic classification and coincides with the position of typical M dwarfs.

The spectrum of the object is plotted in Fig. \ref{fig:spectra}F. No emission lines are seen in the data. In addition to an M3-4 V continuum, excess in the blue part is detected. We have observed the object at several occasions (see Tab. \ref{table:log_obs}) to confirm that the excess is not an observational artefact. This feature is present in all our spectra and seems to be variable to some extent. Slight excess in the blue optical spectral region is also confirmed by the SED of the object (Fig. \ref{fig:seds}F). Such blue excess is often observed in white dwarf-M~dwarf binaries \citep[see e.g., ][and references therein]{2014A&A...570A.107R,2017AJ....154..118S}. However, no trace of broad hydrogen absorption features of a DA white dwarf is detectable in our data. On the other hand, some cool white dwarfs can have a featureless continuum \citep[e.g.,][their Fig. 9, and references therein]{2016ApJ...832...62F}. It seems that a similar excess in the spectra is detected also in some detached M dwarf-M dwarf binaries \citep[e.g.,][especially  19c-3-01405 shows very similar continuum shape as EC 19249-7343]{2012MNRAS.426.1507B}. We should also note that some cataclysmic binaries are located in the similar region of the HR \textit{Gaia} diagram as \mbox{EC 19249-7343} \citep[][]{2020MNRAS.492L..40A}, though, most of the cataclysmic variables exhibit emission lines in their spectra due to the presence of accretion disks \citep[e.g.,][]{2001cvs..book.....H}.

No periodic variability has been detected in the ASAS-SN light curves while there is a scatter of about 0.15 mag in the data. There are two epochs when the star was above the average brightness: at the beginning of the observations of ASAS-SN (at least for 2 months between JD\,2\,455\,978 and JD\,2\,456\,042; 0.3\,mag in the $V$ filter) and during the short (2-5\,days) brightening distinguishable in the light curve between JD\,2\,458\,489 and JD\,2\,458\,491 (0.4 mag in the $g$ filter). Unfortunately, the ZTF data are not available for this object. If the star is indeed a white dwarf-M dwarf binary or a cataclysmic system, the variability on the timescale of hours or days might be detected by careful well-sampled photometric and/or spectroscopic (i.e., variations of radial velocities) observations.

\subsection{V1017 Cyg} The object was classified as a symbiotic star by \citet{2007A&A...469..799S} during the analysis and cataloguing of post-AGB stars. Their classification is based on the star's photometric appearance. Later, the object has also been included in the catalogue of RR Lyrae stars \citep[e.g.][]{2014MNRAS.441..715G}. We should note that the reference to an object called V1017 Cyg appeared in table 2 of \citet{1990ApJ...358..251W}. However, this was only a mistyping of V1016 Cyg, a known symbiotic star, as indicated by the coordinates. Later in the paper (e.g., table 4), they used the correct identifier (V1016 Cyg). No spectrum of V1017~Cyg is available in the literature.

Our optical spectrum of V1017 Cyg (see Fig. \ref{fig:spectra}G) is consistent with a $\sim$\,F8-9 V star. This spectral classification is also confirmed by the multi-frequency SED of the star (Fig. \ref{fig:seds}G). Luminosity class V (see the position of the object in the HR diagram, Fig. \ref{fig:hr}) is confirmed by the absolute magnitude calculated for the distance of $\sim$1\,200\,pc from \textit{Gaia} \citep{2021AJ....161..147B, 2020arXiv201201533G}. The object is the bluest of the whole examined sample (Figs. \ref{fig:hr} and \ref{fig:ir}).

Using the ASAS-SN light curves of the object, we have detected variability with a period of 0.33\,days (Fig. \ref{fig:LCs}G). Combining the photometric variability of the star and the inferred spectral type, we conclude that V1017 Cyg is a W UMa binary (sub-type W).

\subsection{PN K 1-6} The object was discovered by \citet{1962BAICz..13..120K} and classified as a~probable planetary nebula. Later, it was confirmed to be a bona-fide planetary nebula with a variable central star. The appearance and photometric variability were studied in detail by \citet{2011PASA...28...83F}, who suggested that the central star might be a binary or even a triple system. The authors preferred RS CVn or FK Com variable for the classification of the central star, although they also proposed a symbiotic binary.

The obtained optical spectra of the object (one is shown in Fig. \ref{fig:spectra}H) is consistent with a K1-3 V star. The star's positions in the HR and NIR colour-colour diagrams (Figs. \ref{fig:hr}, \ref{fig:ir}), also coincides with dwarf stars of such spectra type. The apparent spectral type seems to change with the brightness of the star which varies with the period of $\sim$\,21\,days and amplitude of $\sim$\,1\,mag in the $V$ filter (Fig. \ref{fig:LCs}H).

The strong fluxes in the UV region detected by \textit{GALEX} (see Fig. \ref{fig:seds}H) could be approximated with the second stellar source with a~temperature of few tens of thousands K. As the object was confirmed to be a planetary nebula \citep{2011PASA...28...83F}, a hot pre-white dwarf or white dwarf is expected. Moreover, several planetary nebulae have binary central stars. The brightness of the object is consistent with the dwarf star (see its position in the HR diagram in Fig. \ref{fig:hr}), if we assume the distance of 258\,pc as inferred by \citet{2021AJ....161..147B} from the \textit{Gaia} EDR3 data \citep{2020arXiv201201533G}. These results, together with the observed $\sim$\,21\,days variability in the light curves of the object, rule out the possible symbiotic classification of PN K1-6.

\subsection{Hen 4-204} This relatively bright, southern object was proposed to be a possible yellow symbiotic system by \citet{2003PASP..115.1367V} on the basis of similarity of the abundance pattern of Hen 4-204 and HD 35155, a known symbiotic binary. Hen 4-204 is classified as a probable extrinsic S-type star \citep[][]{2000A&AS..145...51V}.

The spectrum of the object is plotted in Fig. \ref{fig:spectra}I. The molecular bands of TiO are similar to the ones seen in an M1 III star ($T_{\rm eff} \sim 3\,600\,$K). However, it is clearly seen that there are some differences between the spectra of Hen 4-204 and M1 III star caused by ZrO molecular bands confirming that Hen 4-204 is an S-type star. The optical spectrum of the object is analogous to that of BD~Cam, a~known symbiotic binary with an S3.5/2 cool component \citep[][]{1980ApJS...43..379K}. The symbiotic nature of BD~Cam was confirmed by UV observations showing emission lines with a high ionisation potential \citep[up to 77.5 eV; ][]{1988ESASP.281a.245A}. Unfortunately, spectroscopic observations of Hen 4-204 in this spectral range are not available. The position of the object in the HR diagram (Fig. \ref{fig:hr}) and NIR colour-colour diagram (Fig. \ref{fig:ir}) coincide with the symbiotic stars, and its IR properties satisfy the criteria proposed by \citet[][]{2019MNRAS.483.5077A,2021MNRAS.502.2513A}. However, that cannot be used as a definitive confirmation of the symbiotic nature.

The SED (Fig. \ref{fig:seds}I) suggests an emission excess in the near-UV spectral region compatible with a presence of another stellar object with an effective temperature of $\sim$\,8\,600\,K or a white dwarf with a temperature of $\sim$\,10\,000\,K. If the companion is a~main-sequence star, the cool component of Hen 4-204 would have to be an intrinsic S star. However, the absolute magnitude of the object calculated for a distance of $\sim$1\,320\,pc from \textit{Gaia} \citep{2021AJ....161..147B, 2020arXiv201201533G} is consistent with the~luminosity of a~normal giant star. That means that the cool component of Hen 4-204 is not a thermally-pulsing asymptotic giant branch star, but rather it has been polluted by s-process material from its companion which has evolved faster. Therefore, a~white dwarf is preferred over the main-sequence star to be the hot component of Hen 4-204.

We should note that while the orbital period of 596 days and a~possible pulsation period of 24.8 days was inferred for BD Cam by \citet[][]{1984Obs...104..224G} and \citet[][]{1998A&A...333..952A}, respectively, we have detected the period of 19.12 days in the ASAS-SN $V$ light curve of Hen 4-204 in addition to a bare detection of period $\sim$ 510 days. From all this, we can conclude that Hen 4-204 is an object very similar to BD Cam, but without further observations, especially spectroscopic in the UV region, we cannot definitively classify the system as a~genuine symbiotic binary.

\subsection{V379 Peg} The object has been detected in outburst as a UV-excess star with a blue continuum in the eighties \citep{1981Afz....17..573L}. It was proposed to be either a~cataclysmic variable or symbiotic binary \citep[based on the observations of ][]{1988Afz....28..287K}. However, there is confusion about the position of the object in the sky. Most of the current databases (e.g., SIMBAD, where the star is listed as a confirmed symbiotic binary) point to a red, relatively bright object. The red star was studied by \citet{2002IBVS.5342....1K} who contradicted the symbiotic or cataclysmic nature of the object based on the photometric observations and detected proper motion. On the other hand, \citet{2003IBVS.5368....1H} pointed out that the object originally classified as V379~Peg back in 1981 is not a reddish star, but a~nearby blue object with magnitude $>$\,18\,mag (denoted USNO-A2.0 1125-19982531; see Fig. \ref{fig:v379peg} in the attachment).


We have chosen the red star for our observations as the blue one is too faint for the spectroscopic observations using small telescopes. Although the obtained spectrum has a low S/N ratio (especially in the blue part), it is consistent with M2-3 V star (Fig. \ref{fig:spectra}J). A similar spectral type is inferred from the SED (Fig. \ref{fig:seds}J). The star's position in the NIR colour-colour diagram (Fig. \ref{fig:ir}) coincides with that of early M dwarf stars. This luminosity class is confirmed by the absolute magnitude of the object (see its position in the HR diagram, Fig. \ref{fig:hr}), assuming that V379 Peg is located at the distance of 106\,pc, as given by \textit{Gaia} \citep{2021AJ....161..147B, 2020arXiv201201533G}. Therefore, the object is a dwarf star rather than a symbiotic binary. Spectra with a higher signal to noise ratio may reveal some emission lines due to a chromospheric activity of the star.

The light curves of V379 Peg obtained from the ASAS-SN did not show any significant/periodic variability, but the brightness of the object was at the limit of the survey. In the light curves from the ZTF survey, we have just barely detected the possible period of $\sim$4.3\,days. It is worth mentioning that the object was detected in the GALEX NUV filter on the level higher than expected for a single M dwarf. On the other hand, the star was not detected in the FUV filter. This might be a coincidence with some background source or the star might have a very faint companion of the earlier spectral type. The barely detected variability of $\sim$4.3\,days in ZTF data might suggest the presence of orbital variability in the light curve of V379~Peg.

We should note that the near blue stellar object (USNO-A2.0 1125-19982531), which is probably the one detected in the outburst back in the eighties, still could be a cataclysmic variable. Its parallax in the \textit{Gaia} EDR3 is negative. However, adopting the range of distances 4.8 - 7.6\,kpc obtained from \citet{2021AJ....161..147B}, results in $M_{\rm G}$ = 4.8 - 3.8\,mag. Together with the colour of the object ($G_{\rm BP}$\,-\,$G_{\rm RP}$) = 0.75\,mag, it falls in the region of the \textit{Gaia} HR diagram occupied mainly by nova-like cataclysmic variables \citep[see Fig. 2 of][]{2020MNRAS.492L..40A}. The symbiotic nature is improbable, given its blue colours and low luminosity.



\section{Conclusions}
The proper characterisation of the symbiotic population in the Milky Way and other galaxies, such as the Magellanic Clouds, M\,31 or M\,33 is crucial to understand the evolution of these systems (both pre- and post-symbiotic), the mechanisms responsible for their activity, or their possible contribution to the chemical enrichment of the interstellar matter through the stellar winds, jets, outbursts or even possible supernovae Ia explosions. To allow comprehensive statistical research, we created the New Online Database of Symbiotic Variables. One of the first results achieved using this database is the finding that many of the known symbiotic stars and candidates are only poorly characterised. For this reason, we have initiated the observational spectroscopic campaign with two goals - to confirm or reject the symbiotic nature of the candidates in order to provide a clean sample of symbiotics, and to characterise the poorly studied confirmed systems. Here, we presented the first results of the project.

We have confirmed the symbiotic nature of two objects: V2204~Oph and Hen 3-860. These two objects satisfy the spectroscopic criteria of \citet{2000A&AS..146..407B} for burning symbiotic stars, and also IR criteria presented by \citet{2019MNRAS.483.5077A,2021MNRAS.502.2513A}. Both stars have experienced outbursts during the last hundred years. Hen 3-860 is a~newly discovered southern symbiotic system, representative of an uncommon group of eclipsing symbiotics. Two other objects, Hen 4-204 and V1988 Sgr definitely are not burning symbiotic stars. However, V1988 Sgr still can be an accreting-only symbiotic star and Hen 4-204 seems to be very similar to the known symbiotic star BD Cam with an S-type cool component. Subsequent long-term photometric and spectroscopic observations in various spectral regions would be needed to fully confirm or reject their symbiotic status.

In addition, we have reclassified another six symbiotic candidates as either nearby single main-sequence star (V379 Peg), single giant (IRAS 19050+0001) or binaries (V1017 Cyg, the \mbox{MS-MS} pair; \mbox{PN K1-6}, the system of MS star and white dwarf; EC 19249-7343, possible M dwarf-white dwarf binary or detached M dwarf-M dwarf system; V562 Lyr, appearance of M3 giant with a long-lasting eclipse detected in past).

Candidates for symbiotic novae selected from our New Online Database of Symbiotic Variables \citep{{2019RNAAS...3...28M}} will be the subjects of the second paper of the series.

\section*{Acknowledgements}
We are thankful to an anonymous referee for the comments and suggestions improving the manuscript. This research was supported by the \textit{Charles University}, project GA UK No. 890120, the internal grant VVGS-PF-2021-1746 of the \textit{Faculty of Science, P. J. \v{S}af\'{a}rik University in Ko\v{s}ice}, and the Slovak Research and Development Agency under contract No. APVV-20-0148. The research of MW is supported by the grant GA19-01995S of the Czech Science Foundation.

\section*{Data Availability}

Spectra used in the paper are available in the ARAS Database. Photometric data are accessible from the websites of the surveys. The other data are available on reasonable request to the authors.



\bibliographystyle{mnras}
\bibliography{biblio} 

\begin{thebibliography}{}
\makeatletter
\relax
\def\mn@urlcharsother{\let\do\@makeother \do\$\do\&\do\#\do\^\do\_\do\%\do\~}
\def\mn@doi{\begingroup\mn@urlcharsother \@ifnextchar [ {\mn@doi@}
  {\mn@doi@[]}}
\def\mn@doi@[#1]#2{\def\@tempa{#1}\ifx\@tempa\@empty \href
  {http://dx.doi.org/#2} {doi:#2}\else \href {http://dx.doi.org/#2} {#1}\fi
  \endgroup}
\def\mn@eprint#1#2{\mn@eprint@#1:#2::\@nil}
\def\mn@eprint@arXiv#1{\href {http://arxiv.org/abs/#1} {{\tt arXiv:#1}}}
\def\mn@eprint@dblp#1{\href {http://dblp.uni-trier.de/rec/bibtex/#1.xml}
  {dblp:#1}}
\def\mn@eprint@#1:#2:#3:#4\@nil{\def\@tempa {#1}\def\@tempb {#2}\def\@tempc
  {#3}\ifx \@tempc \@empty \let \@tempc \@tempb \let \@tempb \@tempa \fi \ifx
  \@tempb \@empty \def\@tempb {arXiv}\fi \@ifundefined
  {mn@eprint@\@tempb}{\@tempb:\@tempc}{\expandafter \expandafter \csname
  mn@eprint@\@tempb\endcsname \expandafter{\@tempc}}}

\bibitem[\protect\citeauthoryear{{Abril}, {Schmidtobreick}, {Ederoclite}  \&
  {L{\'o}pez-Sanjuan}}{{Abril} et~al.}{2020}]{2020MNRAS.492L..40A}
{Abril} J.,  {Schmidtobreick} L.,  {Ederoclite} A.,   {L{\'o}pez-Sanjuan} C.,
  2020, \mn@doi [\mnras] {10.1093/mnrasl/slz181}, \href
  {https://ui.adsabs.harvard.edu/abs/2020MNRAS.492L..40A} {492, L40}

\bibitem[\protect\citeauthoryear{{Adelman}}{{Adelman}}{1998}]{1998A&A...333..952A}
{Adelman} S.~J.,  1998, \aap, \href
  {https://ui.adsabs.harvard.edu/abs/1998A&A...333..952A} {333, 952}

\bibitem[\protect\citeauthoryear{{Ake}, {Johnson}  \& {Perry}}{{Ake}
  et~al.}{1988}]{1988ESASP.281a.245A}
{Ake} Thomas~B. I.,  {Johnson} H.~R.,   {Perry} Benjamin~F. J.,  1988, in
  {Longdon} N.,  {Rolfe} E.~J.,  {Kondo} Y.,   {Sahade} J.,  eds,  ESA Special
  Publication Vol. 1, ESA Special Publication. pp 245--248

\bibitem[\protect\citeauthoryear{{Akras}, {Guzman-Ramirez}, {Leal-Ferreira}  \&
  {Ramos-Larios}}{{Akras} et~al.}{2019a}]{2019ApJS..240...21A}
{Akras} S.,  {Guzman-Ramirez} L.,  {Leal-Ferreira} M.~L.,   {Ramos-Larios} G.,
  2019a, \mn@doi [\apjs] {10.3847/1538-4365/aaf88c}, \href
  {https://ui.adsabs.harvard.edu/abs/2019ApJS..240...21A} {240, 21}

\bibitem[\protect\citeauthoryear{{Akras}, {Leal-Ferreira}, {Guzman-Ramirez}  \&
  {Ramos-Larios}}{{Akras} et~al.}{2019b}]{2019MNRAS.483.5077A}
{Akras} S.,  {Leal-Ferreira} M.~L.,  {Guzman-Ramirez} L.,   {Ramos-Larios} G.,
  2019b, \mn@doi [\mnras] {10.1093/mnras/sty3359}, \href
  {https://ui.adsabs.harvard.edu/abs/2019MNRAS.483.5077A} {483, 5077}

\bibitem[\protect\citeauthoryear{{Akras}, {Gon{\c{c}}alves}, {Alvarez-Candal}
  \& {Pereira}}{{Akras} et~al.}{2021}]{2021MNRAS.502.2513A}
{Akras} S.,  {Gon{\c{c}}alves} D.~R.,  {Alvarez-Candal} A.,   {Pereira} C.~B.,
  2021, \mn@doi [\mnras] {10.1093/mnras/stab195}, \href
  {https://ui.adsabs.harvard.edu/abs/2021MNRAS.502.2513A} {502, 2513}

\bibitem[\protect\citeauthoryear{{Allard}}{{Allard}}{2014}]{2014IAUS..299..271A}
{Allard} F.,  2014. pp 271--272, \mn@doi{10.1017/S1743921313008545}

\bibitem[\protect\citeauthoryear{{Allen}}{{Allen}}{1984}]{1984Ap&SS..99..101A}
{Allen} D.~A.,  1984, \mn@doi [\apss] {10.1007/BF00650235}, \href
  {https://ui.adsabs.harvard.edu/\#abs/1984Ap&SS..99..101A} {99, 101}

\bibitem[\protect\citeauthoryear{{Bailer-Jones}, {Rybizki}, {Fouesneau},
  {Demleitner}  \& {Andrae}}{{Bailer-Jones} et~al.}{2021}]{2021AJ....161..147B}
{Bailer-Jones} C.~A.~L.,  {Rybizki} J.,  {Fouesneau} M.,  {Demleitner} M.,
  {Andrae} R.,  2021, \mn@doi [\aj] {10.3847/1538-3881/abd806}, \href
  {https://ui.adsabs.harvard.edu/abs/2021AJ....161..147B} {161, 147}

\bibitem[\protect\citeauthoryear{{Belczy{\'n}ski}, {Miko{\l}ajewska}, {Munari},
  {Ivison}  \& {Friedjung}}{{Belczy{\'n}ski}
  et~al.}{2000}]{2000A&AS..146..407B}
{Belczy{\'n}ski} K.,  {Miko{\l}ajewska} J.,  {Munari} U.,  {Ivison} R.~J.,
  {Friedjung} M.,  2000, \mn@doi [Astronomy and Astrophysics Supplement Series]
  {10.1051/aas:2000280}, \href
  {https://ui.adsabs.harvard.edu/\#abs/2000A&AS..146..407B} {146, 407}

\bibitem[\protect\citeauthoryear{{Bianchi}, {Shiao}  \& {Thilker}}{{Bianchi}
  et~al.}{2017}]{2017ApJS..230...24B}
{Bianchi} L.,  {Shiao} B.,   {Thilker} D.,  2017, \mn@doi [\apjs]
  {10.3847/1538-4365/aa7053}, \href
  {https://ui.adsabs.harvard.edu/abs/2017ApJS..230...24B} {230, 24}

\bibitem[\protect\citeauthoryear{{Birkby} et~al.,}{{Birkby}
  et~al.}{2012}]{2012MNRAS.426.1507B}
{Birkby} J.,  et~al., 2012, \mn@doi [\mnras]
  {10.1111/j.1365-2966.2012.21514.x}, \href
  {https://ui.adsabs.harvard.edu/abs/2012MNRAS.426.1507B} {426, 1507}

\bibitem[\protect\citeauthoryear{{Bond} \& {Kasliwal}}{{Bond} \&
  {Kasliwal}}{2012}]{2012PASP..124.1262B}
{Bond} H.~E.,  {Kasliwal} M.~M.,  2012, \mn@doi [\pasp] {10.1086/668861}, \href
  {https://ui.adsabs.harvard.edu/abs/2012PASP..124.1262B} {124, 1262}

\bibitem[\protect\citeauthoryear{{Cardelli}, {Clayton}  \& {Mathis}}{{Cardelli}
  et~al.}{1989}]{1989ApJ...345..245C}
{Cardelli} J.~A.,  {Clayton} G.~C.,   {Mathis} J.~S.,  1989, \mn@doi [\apj]
  {10.1086/167900}, \href
  {https://ui.adsabs.harvard.edu/abs/1989ApJ...345..245C} {345, 245}

\bibitem[\protect\citeauthoryear{{Corradi} et~al.,}{{Corradi}
  et~al.}{2008}]{2008A&A...480..409C}
{Corradi} R.~L.~M.,  et~al., 2008, \mn@doi [\aap] {10.1051/0004-6361:20078989},
  \href {https://ui.adsabs.harvard.edu/abs/2008A&A...480..409C} {480, 409}

\bibitem[\protect\citeauthoryear{{Dahlmark}}{{Dahlmark}}{2000}]{2000IBVS.4898....1D}
{Dahlmark} L.,  2000, Information Bulletin on Variable Stars, \href
  {https://ui.adsabs.harvard.edu/abs/2000IBVS.4898....1D} {4898, 1}

\bibitem[\protect\citeauthoryear{{Dahm} \& {Simon}}{{Dahm} \&
  {Simon}}{2005}]{2005AJ....129..829D}
{Dahm} S.~E.,  {Simon} T.,  2005, \mn@doi [\aj] {10.1086/426326}, \href
  {https://ui.adsabs.harvard.edu/abs/2005AJ....129..829D} {129, 829}

\bibitem[\protect\citeauthoryear{{Fajardo-Acosta} et~al.,}{{Fajardo-Acosta}
  et~al.}{2016}]{2016ApJ...832...62F}
{Fajardo-Acosta} S.~B.,  et~al., 2016, \mn@doi [\apj]
  {10.3847/0004-637X/832/1/62}, \href
  {https://ui.adsabs.harvard.edu/abs/2016ApJ...832...62F} {832, 62}

\bibitem[\protect\citeauthoryear{{Falc{\'o}n-Barroso},
  {S{\'a}nchez-Bl{\'a}zquez}, {Vazdekis}, {Ricciardelli}, {Cardiel}, {Cenarro},
  {Gorgas}  \& {Peletier}}{{Falc{\'o}n-Barroso}
  et~al.}{2011}]{2011A&A...532A..95F}
{Falc{\'o}n-Barroso} J.,  {S{\'a}nchez-Bl{\'a}zquez} P.,  {Vazdekis} A.,
  {Ricciardelli} E.,  {Cardiel} N.,  {Cenarro} A.~J.,  {Gorgas} J.,
  {Peletier} R.~F.,  2011, \mn@doi [\aap] {10.1051/0004-6361/201116842}, \href
  {https://ui.adsabs.harvard.edu/abs/2011A&A...532A..95F} {532, A95}

\bibitem[\protect\citeauthoryear{{Frew}, {Stanger}, {Fitzgerald}, {Parker},
  {Danaia}  \& et al.}{{Frew} et~al.}{2011}]{2011PASA...28...83F}
{Frew} D.~J.,  {Stanger} J.,  {Fitzgerald} M.,  {Parker} Q.,  {Danaia} L.,   et
  al. 2011, \mn@doi [PASA] {10.1071/AS10017}, \href
  {https://ui.adsabs.harvard.edu/abs/2011PASA...28...83F} {28, 83}

\bibitem[\protect\citeauthoryear{{Frith} et~al.,}{{Frith}
  et~al.}{2013}]{2013MNRAS.435.2161F}
{Frith} J.,  et~al., 2013, \mn@doi [\mnras] {10.1093/mnras/stt1436}, \href
  {https://ui.adsabs.harvard.edu/abs/2013MNRAS.435.2161F} {435, 2161}

\bibitem[\protect\citeauthoryear{{Gaia Collaboration} et~al.,}{{Gaia
  Collaboration} et~al.}{2018}]{2018A&A...616A..10G}
{Gaia Collaboration} et~al., 2018, \mn@doi [\aap]
  {10.1051/0004-6361/201832843}, \href
  {https://ui.adsabs.harvard.edu/abs/2018A&A...616A..10G} {616, A10}

\bibitem[\protect\citeauthoryear{{Gaia Collaboration}, {Brown}, {Vallenari},
  {Prusti}, {de Bruijne}, {Babusiaux}  \& {Biermann}}{{Gaia Collaboration}
  et~al.}{2020}]{2020arXiv201201533G}
{Gaia Collaboration} {Brown} A.~G.~A.,  {Vallenari} A.,  {Prusti} T.,  {de
  Bruijne} J.~H.~J.,  {Babusiaux} C.,   {Biermann} M.,  2020, arXiv e-prints,
  \href {https://ui.adsabs.harvard.edu/abs/2020arXiv201201533G} {p.
  arXiv:2012.01533}

\bibitem[\protect\citeauthoryear{{Gavrilchenko}, {Klein}, {Bloom}  \&
  {Richards}}{{Gavrilchenko} et~al.}{2014}]{2014MNRAS.441..715G}
{Gavrilchenko} T.,  {Klein} C.~R.,  {Bloom} J.~S.,   {Richards} J.~W.,  2014,
  \mn@doi [\mnras] {10.1093/mnras/stu606}, \href
  {https://ui.adsabs.harvard.edu/abs/2014MNRAS.441..715G} {441, 715}

\bibitem[\protect\citeauthoryear{{Griffin}}{{Griffin}}{1984}]{1984Obs...104..224G}
{Griffin} R.~F.,  1984, The Observatory, \href
  {https://ui.adsabs.harvard.edu/abs/1984Obs...104..224G} {104, 224}

\bibitem[\protect\citeauthoryear{{Gromadzki}, {Miko{\l}ajewska}  \&
  {Soszy{\'n}ski}}{{Gromadzki} et~al.}{2013}]{2013AcA....63..405G}
{Gromadzki} M.,  {Miko{\l}ajewska} J.,   {Soszy{\'n}ski} I.,  2013, ACTAA,
  \href {https://ui.adsabs.harvard.edu/abs/2013AcA....63..405G} {63, 405}

\bibitem[\protect\citeauthoryear{{Guilbault}, {Hager}, {Henden}, {Kroll},
  {Kurochkin}, {Moro}  \& {Splittgerber}}{{Guilbault}
  et~al.}{2000}]{2000IBVS.4926....1G}
{Guilbault} P.~R.,  {Hager} T.,  {Henden} A.,  {Kroll} P.,  {Kurochkin} N.~E.,
  {Moro} D.,   {Splittgerber} E.,  2000, Information Bulletin on Variable
  Stars, \href {https://ui.adsabs.harvard.edu/abs/2000IBVS.4926....1G} {4926,
  1}

\bibitem[\protect\citeauthoryear{{Gutierrez-Moreno}, {Moreno}  \&
  {Cortes}}{{Gutierrez-Moreno} et~al.}{1995}]{1995PASP..107..462G}
{Gutierrez-Moreno} A.,  {Moreno} H.,   {Cortes} G.,  1995, \mn@doi [\pasp]
  {10.1086/133575}, \href
  {https://ui.adsabs.harvard.edu/abs/1995PASP..107..462G} {107, 462}

\bibitem[\protect\citeauthoryear{{Hellier}}{{Hellier}}{2001}]{2001cvs..book.....H}
{Hellier} C.,  2001, {Cataclysmic Variable Stars}

\bibitem[\protect\citeauthoryear{{Henden}, {Simonsen}  \& {Sumner}}{{Henden}
  et~al.}{2003}]{2003IBVS.5368....1H}
{Henden} A.,  {Simonsen} M.,   {Sumner} B.,  2003, Information Bulletin on
  Variable Stars, \href {https://ui.adsabs.harvard.edu/abs/2003IBVS.5368....1H}
  {5368, 1}

\bibitem[\protect\citeauthoryear{{Henden}, {Levine}, {Terrell}  \&
  {Welch}}{{Henden} et~al.}{2015}]{2015AAS...22533616H}
{Henden} A.~A.,  {Levine} S.,  {Terrell} D.,   {Welch} D.~L.,  2015, in
  American Astronomical Society Meeting Abstracts \#225. p. 336.16

\bibitem[\protect\citeauthoryear{{Henize}}{{Henize}}{1976}]{1976ApJS...30..491H}
{Henize} K.~G.,  1976, \mn@doi [\apjs] {10.1086/190369}, \href
  {https://ui.adsabs.harvard.edu/abs/1976ApJS...30..491H} {30, 491}

\bibitem[\protect\citeauthoryear{{Hoffleit}}{{Hoffleit}}{1962}]{1962AJ.....67..228H}
{Hoffleit} D.,  1962, \mn@doi [\aj] {10.1086/108699}, \href
  {https://ui.adsabs.harvard.edu/abs/1962AJ.....67..228H} {67, 228}

\bibitem[\protect\citeauthoryear{{I{\l}kiewicz} \&
  {Miko{\l}ajewska}}{{I{\l}kiewicz} \&
  {Miko{\l}ajewska}}{2017}]{2017A&A...606A.110I}
{I{\l}kiewicz} K.,  {Miko{\l}ajewska} J.,  2017, \mn@doi [\aap]
  {10.1051/0004-6361/201731497}, \href
  {https://ui.adsabs.harvard.edu/abs/2017A&A...606A.110I} {606, A110}

\bibitem[\protect\citeauthoryear{{Ilkiewicz}, {Mikolajewska}, {Shara},
  {Udalski}, {Drozd}  \& {Faherty}}{{Ilkiewicz}
  et~al.}{2018}]{2018arXiv181106696I}
{Ilkiewicz} K.,  {Mikolajewska} J.,  {Shara} M.~M.,  {Udalski} A.,  {Drozd} K.,
    {Faherty} J.~K.,  2018, arXiv e-prints, \href
  {https://ui.adsabs.harvard.edu/\#abs/2018arXiv181106696I} {p.
  arXiv:1811.06696}

\bibitem[\protect\citeauthoryear{{Ishihara} et~al.,}{{Ishihara}
  et~al.}{2010}]{2010A&A...514A...1I}
{Ishihara} D.,  et~al., 2010, \mn@doi [\aap] {10.1051/0004-6361/200913811},
  \href {https://ui.adsabs.harvard.edu/abs/2010A&A...514A...1I} {514, A1}

\bibitem[\protect\citeauthoryear{{Kato}, {Yamaoka}, {Torii}, {Ishioka}  \&
  {Uemura}}{{Kato} et~al.}{2002}]{2002IBVS.5342....1K}
{Kato} T.,  {Yamaoka} H.,  {Torii} K.,  {Ishioka} R.,   {Uemura} M.,  2002,
  Information Bulletin on Variable Stars, \href
  {https://ui.adsabs.harvard.edu/abs/2002IBVS.5342....1K} {5342, 1}

\bibitem[\protect\citeauthoryear{{Keenan} \& {Boeshaar}}{{Keenan} \&
  {Boeshaar}}{1980}]{1980ApJS...43..379K}
{Keenan} P.~C.,  {Boeshaar} P.~C.,  1980, \mn@doi [\apjs] {10.1086/190673},
  \href {https://ui.adsabs.harvard.edu/abs/1980ApJS...43..379K} {43, 379}

\bibitem[\protect\citeauthoryear{{Kenyon}}{{Kenyon}}{1986}]{1986syst.book.....K}
{Kenyon} S.~J.,  1986, {The symbiotic stars}.
Cambridge: University Press

\bibitem[\protect\citeauthoryear{{Kenyon} \& {Webbink}}{{Kenyon} \&
  {Webbink}}{1984}]{1984ApJ...279..252K}
{Kenyon} S.~J.,  {Webbink} R.~F.,  1984, \mn@doi [\apj] {10.1086/161888}, \href
  {https://ui.adsabs.harvard.edu/abs/1984ApJ...279..252K} {279, 252}

\bibitem[\protect\citeauthoryear{{Kesseli}, {West}, {Veyette}, {Harrison},
  {Feldman}  \& {Bochanski}}{{Kesseli} et~al.}{2017}]{2017ApJS..230...16K}
{Kesseli} A.~Y.,  {West} A.~A.,  {Veyette} M.,  {Harrison} B.,  {Feldman} D.,
  {Bochanski} J.~J.,  2017, \mn@doi [\apjs] {10.3847/1538-4365/aa656d}, \href
  {https://ui.adsabs.harvard.edu/abs/2017ApJS..230...16K} {230, 16}

\bibitem[\protect\citeauthoryear{{Kochanek}, {Shappee}, {Stanek}, {Holoien},
  {Thompson}  \& et al.}{{Kochanek} et~al.}{2017}]{2017PASP..129j4502K}
{Kochanek} C.~S.,  {Shappee} B.~J.,  {Stanek} K.~Z.,  {Holoien} T.~W.~S.,
  {Thompson} T.~A.,   et al. 2017, \mn@doi [\pasp] {10.1088/1538-3873/aa80d9},
  \href {https://ui.adsabs.harvard.edu/abs/2017PASP..129j4502K} {129, 104502}

\bibitem[\protect\citeauthoryear{{Koester}}{{Koester}}{2010}]{2010MmSAI..81..921K}
{Koester} D.,  2010, \memsai, \href
  {https://ui.adsabs.harvard.edu/abs/2010MmSAI..81..921K} {81, 921}

\bibitem[\protect\citeauthoryear{{Kohoutek}}{{Kohoutek}}{1962}]{1962BAICz..13..120K}
{Kohoutek} L.,  1962, Bulletin of the Astronomical Institutes of
  Czechoslovakia, \href {https://ui.adsabs.harvard.edu/abs/1962BAICz..13..120K}
  {13, 120}

\bibitem[\protect\citeauthoryear{{Kopylov}, {Lipovetsky}, {Somov}, {Somova}  \&
  {Stepanian}}{{Kopylov} et~al.}{1988}]{1988Afz....28..287K}
{Kopylov} I.~M.,  {Lipovetsky} V.~A.,  {Somov} N.~N.,  {Somova} T.~A.,
  {Stepanian} J.~A.,  1988, Astrofizika, \href
  {https://ui.adsabs.harvard.edu/abs/1988Afz....28..287K} {28, 287}

\bibitem[\protect\citeauthoryear{{Lasker} et~al.,}{{Lasker}
  et~al.}{2008}]{2008AJ....136..735L}
{Lasker} B.~M.,  et~al., 2008, \mn@doi [\aj] {10.1088/0004-6256/136/2/735},
  \href {https://ui.adsabs.harvard.edu/abs/2008AJ....136..735L} {136, 735}

\bibitem[\protect\citeauthoryear{{L{\'e}pine} \& {Gaidos}}{{L{\'e}pine} \&
  {Gaidos}}{2011}]{2011AJ....142..138L}
{L{\'e}pine} S.,  {Gaidos} E.,  2011, \mn@doi [\aj]
  {10.1088/0004-6256/142/4/138}, \href
  {https://ui.adsabs.harvard.edu/abs/2011AJ....142..138L} {142, 138}

\bibitem[\protect\citeauthoryear{{Lipovetsky} \& {Stepanian}}{{Lipovetsky} \&
  {Stepanian}}{1981}]{1981Afz....17..573L}
{Lipovetsky} V.~A.,  {Stepanian} J.~A.,  1981, Astrofizika, \href
  {https://ui.adsabs.harvard.edu/abs/1981Afz....17..573L} {17, 573}

\bibitem[\protect\citeauthoryear{{Luna}, {Sokoloski}, {Mukai}  \&
  {Nelson}}{{Luna} et~al.}{2013}]{2013A&A...559A...6L}
{Luna} G.~J.~M.,  {Sokoloski} J.~L.,  {Mukai} K.,   {Nelson} T.,  2013, \mn@doi
  [\aap] {10.1051/0004-6361/201220792}, \href
  {https://ui.adsabs.harvard.edu/abs/2013A&A...559A...6L} {559, A6}

\bibitem[\protect\citeauthoryear{{Malkov}, {Kovaleva}, {Sichevsky}  \&
  {Zhao}}{{Malkov} et~al.}{2020}]{2020RAA....20..139M}
{Malkov} O.,  {Kovaleva} D.,  {Sichevsky} S.,   {Zhao} G.,  2020, \mn@doi
  [Research in Astronomy and Astrophysics] {10.1088/1674-4527/20/9/139}, \href
  {https://ui.adsabs.harvard.edu/abs/2020RAA....20..139M} {20, 139}

\bibitem[\protect\citeauthoryear{{Masci} et~al.,}{{Masci}
  et~al.}{2019}]{2019PASP..131a8003M}
{Masci} F.~J.,  et~al., 2019, \mn@doi [\pasp] {10.1088/1538-3873/aae8ac}, \href
  {https://ui.adsabs.harvard.edu/abs/2019PASP..131a8003M} {131, 018003}

\bibitem[\protect\citeauthoryear{{Merc}, {G{\'a}lis}  \& {Wolf}}{{Merc}
  et~al.}{2019}]{2019RNAAS...3...28M}
{Merc} J.,  {G{\'a}lis} R.,   {Wolf} M.,  2019, \mn@doi [Research Notes of the
  American Astronomical Society] {10.3847/2515-5172/ab0429}, \href
  {https://ui.adsabs.harvard.edu/abs/2019RNAAS...3...28M} {3, 28}

\bibitem[\protect\citeauthoryear{{Merc}, {G{\'a}lis}, {K{\'a}ra}, {Wolf}  \&
  {Vra{\v{s}}{\v{t}}{\'a}k}}{{Merc} et~al.}{2020}]{2020MNRAS.499.2116M}
{Merc} J.,  {G{\'a}lis} R.,  {K{\'a}ra} J.,  {Wolf} M.,
  {Vra{\v{s}}{\v{t}}{\'a}k} M.,  2020, \mn@doi [\mnras]
  {10.1093/mnras/staa3063}, \href
  {https://ui.adsabs.harvard.edu/abs/2020MNRAS.499.2116M} {499, 2116}

\bibitem[\protect\citeauthoryear{{Miko{\l}ajewska}}{{Miko{\l}ajewska}}{2012}]{2012BaltA..21....5M}
{Miko{\l}ajewska} J.,  2012, \mn@doi [Baltic Astronomy]
  {10.1515/astro-2017-0352}, \href
  {https://ui.adsabs.harvard.edu/\#abs/2012BaltA..21....5M} {21, 5}

\bibitem[\protect\citeauthoryear{{Miszalski} \& {Miko{\l}ajewska}}{{Miszalski}
  \& {Miko{\l}ajewska}}{2014}]{2014MNRAS.440.1410M}
{Miszalski} B.,  {Miko{\l}ajewska} J.,  2014, \mn@doi [\mnras]
  {10.1093/mnras/stu292}, \href
  {https://ui.adsabs.harvard.edu/\#abs/2014MNRAS.440.1410M} {440, 1410}

\bibitem[\protect\citeauthoryear{{Mukai} et~al.,}{{Mukai}
  et~al.}{2016}]{2016MNRAS.461L...1M}
{Mukai} K.,  et~al., 2016, \mn@doi [\mnras] {10.1093/mnrasl/slw087}, \href
  {https://ui.adsabs.harvard.edu/abs/2016MNRAS.461L...1M} {461, L1}

\bibitem[\protect\citeauthoryear{{Munari}}{{Munari}}{2019}]{2019arXiv190901389M}
{Munari} U.,  2019, arXiv e-prints, \href
  {https://ui.adsabs.harvard.edu/abs/2019arXiv190901389M} {p. arXiv:1909.01389}

\bibitem[\protect\citeauthoryear{{Munari} et~al.,}{{Munari}
  et~al.}{2021}]{2021arXiv210402686M}
{Munari} U.,  et~al., 2021, arXiv e-prints, \href
  {https://ui.adsabs.harvard.edu/abs/2021arXiv210402686M} {p. arXiv:2104.02686}

\bibitem[\protect\citeauthoryear{{Murset} \& {Nussbaumer}}{{Murset} \&
  {Nussbaumer}}{1994}]{1994A&A...282..586M}
{Murset} U.,  {Nussbaumer} H.,  1994, \aap, \href
  {https://ui.adsabs.harvard.edu/abs/1994A&A...282..586M} {282, 586}

\bibitem[\protect\citeauthoryear{{M{\"u}rset} \& {Schmid}}{{M{\"u}rset} \&
  {Schmid}}{1999}]{1999A&AS..137..473M}
{M{\"u}rset} U.,  {Schmid} H.~M.,  1999, \mn@doi [Astronomy and Astrophysics
  Supplement Series] {10.1051/aas:1999105}, \href
  {https://ui.adsabs.harvard.edu/abs/1999A&AS..137..473M} {137, 473}

\bibitem[\protect\citeauthoryear{{Neugebauer} \& {Leighton}}{{Neugebauer} \&
  {Leighton}}{1969}]{1969tmss.book.....N}
{Neugebauer} G.,  {Leighton} R.~B.,  1969, {Two-micron sky survey. A
  preliminary catalogue}.
Washington: NASA

\bibitem[\protect\citeauthoryear{{O'Donoghue}, {Kilkenny}, {Koen}, {Hambly},
  {MacGillivray}  \& {Stobie}}{{O'Donoghue} et~al.}{2013}]{2013MNRAS.431..240O}
{O'Donoghue} D.,  {Kilkenny} D.,  {Koen} C.,  {Hambly} N.,  {MacGillivray} H.,
   {Stobie} R.~S.,  2013, \mn@doi [\mnras] {10.1093/mnras/stt158}, \href
  {https://ui.adsabs.harvard.edu/abs/2013MNRAS.431..240O} {431, 240}

\bibitem[\protect\citeauthoryear{{Phillips}}{{Phillips}}{2007}]{2007MNRAS.376.1120P}
{Phillips} J.~P.,  2007, \mn@doi [\mnras] {10.1111/j.1365-2966.2007.11484.x},
  \href {https://ui.adsabs.harvard.edu/abs/2007MNRAS.376.1120P} {376, 1120}

\bibitem[\protect\citeauthoryear{{Pojmanski}}{{Pojmanski}}{1997}]{1997AcA....47..467P}
{Pojmanski} G.,  1997, \actaa, \href
  {https://ui.adsabs.harvard.edu/abs/1997AcA....47..467P} {47, 467}

\bibitem[\protect\citeauthoryear{{Ramos-Larios} \& {Phillips}}{{Ramos-Larios}
  \& {Phillips}}{2005}]{2005MNRAS.357..732R}
{Ramos-Larios} G.,  {Phillips} J.~P.,  2005, \mn@doi [\mnras]
  {10.1111/j.1365-2966.2005.08713.x}, \href
  {https://ui.adsabs.harvard.edu/abs/2005MNRAS.357..732R} {357, 732}

\bibitem[\protect\citeauthoryear{{Ren} et~al.,}{{Ren}
  et~al.}{2014}]{2014A&A...570A.107R}
{Ren} J.~J.,  et~al., 2014, \mn@doi [\aap] {10.1051/0004-6361/201423689}, \href
  {https://ui.adsabs.harvard.edu/abs/2014A&A...570A.107R} {570, A107}

\bibitem[\protect\citeauthoryear{{Rodr{\'\i}guez-Flores}, {Corradi}, {Mampaso},
  {Garc{\'\i}a-Alvarez}, {Munari}, {Greimel}, {Rubio-D{\'\i}ez}  \& {Santand
  er-Garc{\'\i}a}}{{Rodr{\'\i}guez-Flores} et~al.}{2014}]{2014A&A...567A..49R}
{Rodr{\'\i}guez-Flores} E.~R.,  {Corradi} R.~L.~M.,  {Mampaso} A.,
  {Garc{\'\i}a-Alvarez} D.,  {Munari} U.,  {Greimel} R.,  {Rubio-D{\'\i}ez}
  M.~M.,   {Santand er-Garc{\'\i}a} M.,  2014, \mn@doi [\aap]
  {10.1051/0004-6361/201323182}, \href
  {https://ui.adsabs.harvard.edu/abs/2014A&A...567A..49R} {567, A49}

\bibitem[\protect\citeauthoryear{{Ross}}{{Ross}}{1926}]{1926AJ.....36..122R}
{Ross} F.~E.,  1926, \mn@doi [\aj] {10.1086/104698}, \href
  {https://ui.adsabs.harvard.edu/abs/1926AJ.....36..122R} {36, 122}

\bibitem[\protect\citeauthoryear{{Samus'}}{{Samus'}}{1983}]{1983MitVS...9...87S}
{Samus'} N.~N.,  1983, Zentralinstitut fuer Astrophysik Sternwarte Sonneberg
  Mitteilungen ueber Veraenderliche Sterne, \href
  {https://ui.adsabs.harvard.edu/abs/1983MitVS...9...87S} {9, 87}

\bibitem[\protect\citeauthoryear{{Samus'}, {Kazarovets}, {Durlevich}, {Kireeva}
   \& {Pastukhova}}{{Samus'} et~al.}{2017}]{2017ARep...61...80S}
{Samus'} N.~N.,  {Kazarovets} E.~V.,  {Durlevich} O.~V.,  {Kireeva} N.~N.,
  {Pastukhova} E.~N.,  2017, \mn@doi [Astronomy Reports]
  {10.1134/S1063772917010085}, \href
  {https://ui.adsabs.harvard.edu/abs/2017ARep...61...80S} {61, 80}

\bibitem[\protect\citeauthoryear{{Schlafly} \& {Finkbeiner}}{{Schlafly} \&
  {Finkbeiner}}{2011}]{2011ApJ...737..103S}
{Schlafly} E.~F.,  {Finkbeiner} D.~P.,  2011, \mn@doi [\apj]
  {10.1088/0004-637X/737/2/103}, \href
  {https://ui.adsabs.harvard.edu/abs/2011ApJ...737..103S} {737, 103}

\bibitem[\protect\citeauthoryear{{Schmid}}{{Schmid}}{1989}]{1989A&A...211L..31S}
{Schmid} H.~M.,  1989, \aap, \href
  {https://ui.adsabs.harvard.edu/abs/1989A&A...211L..31S} {211, L31}

\bibitem[\protect\citeauthoryear{{Shappee}, {Prieto}, {Grupe}, {Kochanek},
  {Stanek}  \& et al.}{{Shappee} et~al.}{2014}]{2014ApJ...788...48S}
{Shappee} B.~J.,  {Prieto} J.~L.,  {Grupe} D.,  {Kochanek} C.~S.,  {Stanek}
  K.~Z.,   et al. 2014, \mn@doi [\apj] {10.1088/0004-637X/788/1/48}, \href
  {https://ui.adsabs.harvard.edu/abs/2014ApJ...788...48S} {788, 48}

\bibitem[\protect\citeauthoryear{{Skinner}, {Morgan}, {West}, {L{\'e}pine}  \&
  {Thorstensen}}{{Skinner} et~al.}{2017}]{2017AJ....154..118S}
{Skinner} J.~N.,  {Morgan} D.~P.,  {West} A.~A.,  {L{\'e}pine} S.,
  {Thorstensen} J.~R.,  2017, \mn@doi [\aj] {10.3847/1538-3881/aa83b5}, \href
  {https://ui.adsabs.harvard.edu/abs/2017AJ....154..118S} {154, 118}

\bibitem[\protect\citeauthoryear{{Skopal}}{{Skopal}}{2005}]{2005A&A...440..995S}
{Skopal} A.,  2005, \mn@doi [\aap] {10.1051/0004-6361:20034262}, \href
  {https://ui.adsabs.harvard.edu/abs/2005A&A...440..995S} {440, 995}

\bibitem[\protect\citeauthoryear{{Skrutskie} et~al.,}{{Skrutskie}
  et~al.}{2006}]{2006AJ....131.1163S}
{Skrutskie} M.~F.,  et~al., 2006, \mn@doi [\aj] {10.1086/498708}, \href
  {https://ui.adsabs.harvard.edu/abs/2006AJ....131.1163S} {131, 1163}

\bibitem[\protect\citeauthoryear{{Szczerba}, {Si{\'o}dmiak}, {Stasi{\'n}ska}
  \& {Borkowski}}{{Szczerba} et~al.}{2007}]{2007A&A...469..799S}
{Szczerba} R.,  {Si{\'o}dmiak} N.,  {Stasi{\'n}ska} G.,   {Borkowski} J.,
  2007, \mn@doi [\aap] {10.1051/0004-6361:20067035}, \href
  {https://ui.adsabs.harvard.edu/abs/2007A&A...469..799S} {469, 799}

\bibitem[\protect\citeauthoryear{{Teyssier}}{{Teyssier}}{2019}]{2019CoSka..49..217T}
{Teyssier} F.,  2019, Contributions of the Astronomical Observatory Skalnate
  Pleso, \href {https://ui.adsabs.harvard.edu/abs/2019CoSka..49..217T} {49,
  217}

\bibitem[\protect\citeauthoryear{{Van Eck}, {Jorissen}, {Udry}, {Mayor},
  {Burki}, {Burnet}  \& {Catchpole}}{{Van Eck}
  et~al.}{2000}]{2000A&AS..145...51V}
{Van Eck} S.,  {Jorissen} A.,  {Udry} S.,  {Mayor} M.,  {Burki} G.,  {Burnet}
  M.,   {Catchpole} R.,  2000, \mn@doi [\aaps] {10.1051/aas:2000349}, \href
  {https://ui.adsabs.harvard.edu/abs/2000A&AS..145...51V} {145, 51}

\bibitem[\protect\citeauthoryear{{Vanture} \& {Wallerstein}}{{Vanture} \&
  {Wallerstein}}{2003}]{2003PASP..115.1367V}
{Vanture} A.~D.,  {Wallerstein} G.,  2003, \mn@doi [\pasp] {10.1086/380583},
  \href {https://ui.adsabs.harvard.edu/abs/2003PASP..115.1367V} {115, 1367}

\bibitem[\protect\citeauthoryear{{Wannier}, {Sahai}, {Andersson}  \&
  {Johnson}}{{Wannier} et~al.}{1990}]{1990ApJ...358..251W}
{Wannier} P.~G.,  {Sahai} R.,  {Andersson} B.~G.,   {Johnson} H.~R.,  1990,
  \mn@doi [\apj] {10.1086/168980}, \href
  {https://ui.adsabs.harvard.edu/abs/1990ApJ...358..251W} {358, 251}

\bibitem[\protect\citeauthoryear{{Whitelock}, {Feast}  \& {Van
  Leeuwen}}{{Whitelock} et~al.}{2008}]{2008MNRAS.386..313W}
{Whitelock} P.~A.,  {Feast} M.~W.,   {Van Leeuwen} F.,  2008, \mn@doi [\mnras]
  {10.1111/j.1365-2966.2008.13032.x}, \href
  {https://ui.adsabs.harvard.edu/abs/2008MNRAS.386..313W} {386, 313}

\bibitem[\protect\citeauthoryear{{Wolf} et~al.,}{{Wolf}
  et~al.}{2018}]{2018PASA...35...10W}
{Wolf} C.,  et~al., 2018, \mn@doi [\pasa] {10.1017/pasa.2018.5}, \href
  {https://ui.adsabs.harvard.edu/abs/2018PASA...35...10W} {35, e010}

\bibitem[\protect\citeauthoryear{{Wray}}{{Wray}}{1966}]{1966PhDT.........3W}
{Wray} J.~D.,  1966, PhD thesis, NORTHWESTERN UNIVERSITY.

\bibitem[\protect\citeauthoryear{{Wright} et~al.,}{{Wright}
  et~al.}{2010}]{2010AJ....140.1868W}
{Wright} E.~L.,  et~al., 2010, \mn@doi [\aj] {10.1088/0004-6256/140/6/1868},
  \href {https://ui.adsabs.harvard.edu/abs/2010AJ....140.1868W} {140, 1868}

\bibitem[\protect\citeauthoryear{{Zhang}, {Chen}  \& {Yang}}{{Zhang}
  et~al.}{2005}]{2005NewA...10..325Z}
{Zhang} P.,  {Chen} P.~S.,   {Yang} H.~T.,  2005, \mn@doi [\na]
  {10.1016/j.newast.2004.12.002}, \href
  {https://ui.adsabs.harvard.edu/abs/2005NewA...10..325Z} {10, 325}

\makeatother
\end{thebibliography}



\newpage
\appendix

\begin{table}\section{Log of observations}
\caption{Log of observations. Star symbol (*) denote the spectra used in Fig. \ref{fig:spectra}. Observer codes: BHQ = T. Bohlsen, HBO = H. Boussier, CBO = C. Boussin, BUI = C. Buil, PCA = P. Cazzato, IBR = I. Diarrasouba, FAS = F. Sims, VLZ = P. Velez, MAV = Martin Vrašťák.}             
\label{table:log_obs}      
\centering
\begin{tabular}{lllcc}
\hline\hline
Object & JD& Res. & $\lambda_{\rm min}$-$\lambda_{\rm max}$ & Obs. \\
 & 2\,459\,.. &  & [\AA] &  \\\hline
*Hen 3-860 & 049.438 & 1176 & 3800-7501 & VLZ \\
Hen 3-860 & 051.524 & 1778 & 4000-8000 & BHQ \\
Hen 3-860 & 352.935 & 1271 & 5508-7549 & VLZ \\
Hen 3-860 & 353.951 & 1068 & 3924-5966 & VLZ \\
Hen 3-860 & 355.939 & 1031 & 3920-5960 & VLZ \\
Hen 3-860 & 356.892 & 1055 & 3920-5969 & VLZ \\
Hen 3-860 & 357.065 & 1014 & 3920-5955 & VLZ \\
Hen 3-860 & 358.928 & 1271 & 5560-7622 & VLZ \\
V2204 Oph & 060.887 & 782 & 3871-7380 & HBO \\
*V2204 Oph & 066.460 & 951 & 3851-7501 & VLZ \\
V1988 Sgr & 125.438 & 1126 & 3801-7401 & VLZ \\
*V1988 Sgr & 347.116 & 1017 & 3765 - 5752 & VLZ \\
*V1988 Sgr & 351.048 & 1272 & 5524 - 7564 & VLZ \\
LP Sgr & 347.144 & 1021 & 3765 - 5752 & VLZ \\
LP Sgr & 351.089 & 1251 & 5523 - 7569 & VLZ \\
V562 Lyr & 042.014 & 564 & 3550-6027 & IBR \\
*V562 Lyr & 049.054 & 506 & 3800-7396 & CBO \\
*IRAS 19050+0001 & 058.355 & 1061 & 4000-7276 & FAS \\
EC 19249-7343 & 060.412 & 895 & 3801-7500 & VLZ \\
EC 19249-7343 & 060.469 & 1623 & 3800-8000 & BHQ \\
EC 19249-7343 & 352.026 & 1267 & 5518-7560 & VLZ \\
EC 19249-7343 & 354.013 & 1011 & 3930-5968 & VLZ \\
*EC 19249-7343 & 360.018 & 1547 & 3900-8199 & BHQ \\
*V1017 Cyg & 040.013 & 428 & 3700-6799 & BUI \\
V1017 Cyg & 040.251 & 1032 & 4001-7276 & FAS \\
PN K1-6 & 039.304 & 1062 & 3825-7249 & FAS \\
PN K1-6 & 039.916 & 432 & 3700-6799 & BUI \\
PN K1-6 & 051.002 & 538 & 3802-7300 & PCA \\
*PN K1-6 & 069.505 & 630 & 3400-8499 & MAV \\
*Hen 4-204 & 357.297 & 1084 & 3907-5978 & VLZ \\
*Hen 4-204 & 359.276 & 1261 & 5582-7612 & VLZ \\
*V379 Peg & 039.095 & 381 & 3745-6797 & BUI\\\hline
\end{tabular}
\end{table}
\pagebreak

\begin{figure}\section{Finding charts of selected objects}
\centering
\includegraphics[width=\columnwidth]{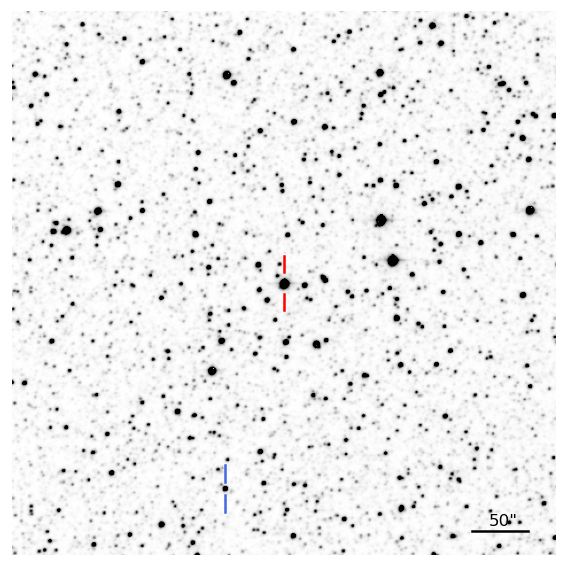}
\caption{Field of V1988 Sgr from the ZTF survey (filter \textit{r}, 480" x 480"). The studied star is marked red, LP Sgr discussed in the text is shown with blue marks.}
\label{fig:v1988sgr}
\end{figure}

\begin{figure}
\centering
\includegraphics[width=\columnwidth]{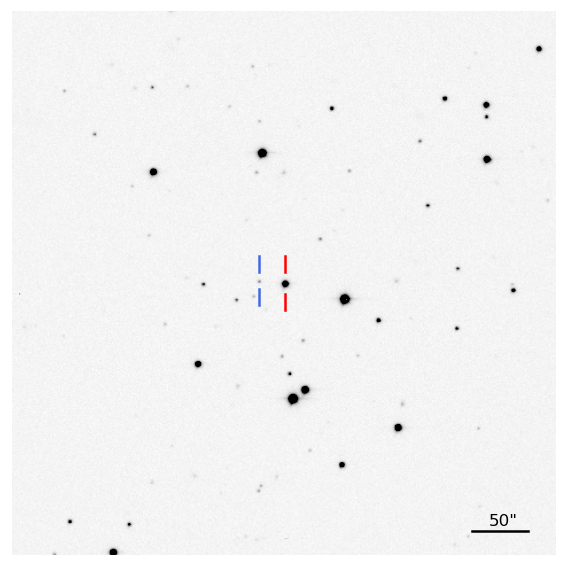}
\caption{Field of V379 Peg from the ZTF survey (filter \textit{r}, 480" x 480"). The studied star is marked red, star USNO-A2.0 1125-19982531 discussed in the text is shown with blue marks.}
\label{fig:v379peg}
\end{figure}

\bsp	
\label{lastpage}
\end{document}